\documentclass[12pt]{article}
\usepackage{epsfig}
\usepackage{a4}
\usepackage{latexsym}
\usepackage{cite}

\usepackage{pslatex}
\usepackage[latin1]{inputenc}
\usepackage[T1]{fontenc}

\usepackage{color,colordvi}
\def\colour4colour#1{\Blue{#1}}

\def\frak#1#2{{\textstyle{{#1}\over{#2}}}}

\newcommand{\beq}{\begin{equation}}
\newcommand{\eeq}{\end{equation}}
\newcommand{\bea}{\begin{eqnarray}}
\newcommand{\eea}{\end{eqnarray}}
\newcommand{\nn}{\nonumber}
\newcommand{\nin}{\noindent}
\newcommand{\ra}{\rightarrow}
\newcommand{\hspn}{{\hspace{-3mm}}}
\newcommand{\DD}{{\cal D}}

\newcommand{\MSb}{$\overline{\mbox{MS}}$}
\newcommand{\als}{\alpha_{\rm s}}
\newcommand{\as}{a_{\rm s}}
\newcommand{\Qs}{Q^{\:\! 2}}
\def\Fone{{F_{\:\! 1}}}
\def\Ftwo{{F_{\:\! 2}}}

\def\FT{{F_{\:\! T}}}
\def\x1{{(1 \! - \! x)}}

\begin{document}
\setlength{\parskip}{0.2cm}
\setlength{\baselineskip}{0.525cm}

\def\z#1{{\zeta_{#1}}}
\def\zs2{{\zeta_{2}^{\,2}}}
\def\ca{{C^{}_A}}
\def\cas{{C^{\, 2}_A}}
\def\cat{{C^{\, 3}_A}}
\def\cf{{C^{}_F}}
\def\cfs{{C^{\, 2}_F}}
\def\cft{{C^{\, 3}_F}}
\def\nf{{n^{}_{\! f}}}
\def\n2f{{n^{\,2}_{\! f}}}
\def\nfs{{n^{\,2}_{\! f}}}
\def\nft{{n^{\,3}_{\! f}}}
\def\caf{{C^{}_{\!AF}}}
\def\cafs{{C^{\,2}_{\!AF}}}
\def\b#1{{{\beta}_{#1}}}
\def\bs#1{{{\beta}_{#1}^{\,2}}}

\def\pqq(#1){p_{\rm{qq}}(#1)}
\def\pqg(#1){p_{\rm{qg}}(#1)}
\def\pgq(#1){p_{\rm{gq}}(#1)}
\def\pgg(#1){p_{\rm{gg}}(#1)}
\def\H(#1){{\rm{H}}_{#1}}
\def\Hh(#1,#2){{\rm{H}}_{#1,#2}}
\def\Hhh(#1,#2,#3){{\rm{H}}_{#1,#2,#3}}
\def\Hhhh(#1,#2,#3,#4){{\rm{H}}_{#1,#2,#3,#4}}

\begin{titlepage}
\noindent
LTH 916 \hspace*{\fill} July 2011\\[0.5mm] 
DESY 11-108\\[0.5mm] 
SFB/CPP-11-33 \\[0.5mm] 
LPN11-34 \\
\vspace{1.2cm}
\begin{center}
\Large
{\bf On the Next-to-Next-to-Leading Order Evolution } \\[2mm]
{\bf of Flavour-Singlet Fragmentation Functions} \\
\vspace{2.0cm}
\large
A.A. Almasy$^{\, a}$, S. Moch$^{\, b}$ and A. Vogt$^{\, a}$\\
\vspace{1.5cm}
\normalsize
{\it $^a$Department of Mathematical Sciences, University of Liverpool \\
\vspace{0.1cm}
Liverpool L69 3BX, United Kingdom}\\[5mm]
{\it $^b$Deutsches Elektronensynchrotron DESY \\
\vspace{0.1cm}
Platanenallee 6, D--15738 Zeuthen, Germany}\\[2.5cm]
\vfill
\large
{\bf Abstract}
\vspace{-0.2cm}
\end{center}
We present the third-order contributions to the quark-gluon and gluon-quark 
timelike splitting functions for the evolution of fragmentation functions in
perturbative QCD. These quantities have been derived by studying physical 
evolution kernels for photon- and Higgs-exchange structure functions in 
deep-inelastic scattering and their counterparts in semi-inclusive 
annihilation, together with constraints from the momentum sum rule and the
supersymmetric limit. For this purpose we have also calculated the second-order
coefficient functions for one-hadron inclusive Higgs decay in the heavy-top
limit. A numerically tolerable uncertainty remains for the quark-gluon 
splitting function, which does not affect the endpoint logarithms for small and
large momentum fractions. We briefly discuss these limits and illustrate the 
numerical impact of the third-order corrections.
Compact and accurate parametrizations are provided for all third-order timelike
splitting functions.
\vspace{0.5cm}
\end{titlepage}
%
% -----------------------------------------------------------------------------
%
\nin
In this article we address the scale dependence (evolution) of the parton
fragmentation distributions (functions) $\,D_{\! f}^{\,h}(x,\Qs)$, see 
Ref.~\cite{PDG10} for an introductory overview, at the next-to-next-to-leading 
\mbox{order} (NNLO) in massless perturbative QCD. 
Here $x$ represents the fractional momentum of the final-state parton $f$ 
transferred to the outgoing hadron $h$; and $\Qs$ is a (timelike) hard scale, 
for instance the squared momentum of the virtual photon or $Z$-boson in 
semi-inclusive electron-positron annihilation (SIA), $\, e^+ e^- \ra\, \gamma
\,, \:Z \,\ra\, h + X$, where $X$ stands for all accessible hadronic states.
 
The evolution of the fragmentation distributions is given by
\beq
\label{eq:Devol}
  {d \over d \ln \Qs} \; D_{a}^{\,h} (x,\Qs) \:\: = \:\:
  \int_x^1 {dz \over z} \; P^{\,T}_{ba} \left( z,\als (\Qs) \right)
  \:  D_{b}^{\,h} \Big(\, {x \over z},\, \Qs \Big) \; ,
\eeq
where the summation over $\,b \, =\, q_i^{},\:\bar{q}_i^{},\:g\,$ for 
$i = 1,\,\ldots,\, \nf\,$ is understood, and $\nf$ denotes the number of 
effectively massless quark flavours. 
Unlike the functions $\,D_{\! f}^{\,h}(x,\Qs)$, the `timelike' splitting 
functions $\,P^{\,T}_{ba}\,$ can be expanded in powers of the strong coupling 
$\als$,
\beq
\label{eq:PTexp}
  P^{\,T}_{ba} \left( x,\als (\Qs) \right) \:\: = \:\:
  \as \, P_{ba}^{(0)T}(x) \: + \: \as^{\:\!2} \, P_{ba}^{(1)T}(x)
  \: +\: \as^{\:\!3}\, P_{ba}^{(2)T}(x) \: +\: \ldots \;\: .
\eeq 
We normalize the expansion parameter as $\,\as = \als/ (4\pi)$ and use, without
loss of information, the standard \MSb\ scheme with the choice 
$\mu_r^{\,2} = \mu_{\!f}^{\,2} = \Qs$ for the renormalization and 
fragmentation (final-state mass factorization) scale.
The system (\ref{eq:Devol}) of $(2\nf+\!1)\!\times\!(2\nf+\!1)$ coupled 
equations~can be decomposed into $2\nf+\!1$ scalar flavour non-singlet 
equations and the $2\!\times\!2$ flavour-singlet system
\beq
\label{eq:Dsgevol}
  \frac{d}{d \ln \Qs}
  \left( \begin{array}{c} \!D^{}_{\rm q}\! \\ \!D_{\rm g}\! \end{array} \right)
  \: = \: \left( \begin{array}{cc}
         P_{\rm qq}^{\,T} & P_{\rm gq}^{\,T} \\[2mm]
         P_{\rm qg}^{\,T} & P_{\rm gg}^{\,T}
   \end{array} \right) \otimes
  \left( \begin{array}{c} \!D^{}_{\rm q}\! \\ \!D_{\rm g}\! \end{array} \right)
  \quad\: \mbox{with} \quad\:
  D^{}_{\rm q} \; = \; \sum_{i=1}^{\nf} ( D_{q_i^{}} + D_{\bar{q}_i^{}} ) \:\: .
\eeq
Here $\otimes$ abbreviates the convolution in Eq.~(\ref {eq:Devol}), and we have
suppressed all functional dependences. 

The leading-order (LO) splitting functions $P^{(0)T\!}(x)$ \cite{GP78} are 
identical to their `spacelike' counterparts \cite{AP77} for the evolution of 
the initial-state parton distributions (where the matrix in 
Eq.~(\ref{eq:Dsgevol}) is transposed), a fact often referred to as the
Gribov-Lipatov relation \cite{GL72}. 
The next-to-leading order contributions $P^{(1)T}$ were derived by several 
groups about thirty years ago \cite{CFP80,FP80,KKST80,FKL81,MOKK01}. 
Unlike the spacelike case \cite{MVV3,MVV4}, where the calculations can be 
performed via forward scattering amplitudes, the NNLO corrections $P^{(2)T}$ 
have eluded a direct calculation in terms of Feynman diagrams so far. 

The (three) non-singlet quantities $P_{\rm ns}^{(2)T}$ and the diagonal entries
in Eq.~(\ref{eq:Dsgevol}) have been determined by two of us a couple of years
ago \cite{MMV06,MV2} via analytic continuations (see below) of the unfactorized
partonic structure functions from the spacelike (deep-inelastic scattering, 
DIS) to the timelike SIA case, supplemented by complementary considerations
based on Ref.~\cite{DMS05}. 
Only the second Mellin moments of $P_{\rm gq}^{(2)T\!}(x)$ and 
$P_{\rm qg}^{(2)T\!}(x)$ are fully known at this point \cite{MV2}, since these 
quantities are fixed by the momentum sum rule and the diagonal entries.

Beyond the leading order, there is no direct relation between the spacelike 
splitting functions, or their $x \ra 1/x$ analytic continuations ({\it AC}), 
and their timelike counterparts in the \MSb\ scheme. 
However such a relation exists for spacelike and timelike physical evolution 
kernels $K(x,\als)$ for photon-exchange DIS and SIA structure functions at NLO 
\cite{BRvN00}, see also Ref.~\cite{SV96}.
Defining the expansion coefficients $K^{(n)}$ as in Eq.~(\ref{eq:PTexp}), it
can schematically be written as
\beq
\label{eq:ACphys1}
  \mbox{\it AC}\: \big[ K^{(n)S}(x) \big] \;=\; K^{(n)T}(x)
  \quad\: \mbox{for} \quad\: n\:=\: 0,\: 1 \:\: .
\eeq

In order to access all four splitting functions, we study the physical 
evolution kernels for the system $\Fone$ and $F_\phi$ of flavour-singlet DIS 
structure function.
The former quantity is chosen, instead of $\Ftwo$ in Ref.~\cite{SMVV1}, since 
it directly corresponds to the transverse fragmentation function $\FT$ in SIA.
The NNLO coefficient functions for these observables are known from 
Refs.~\cite{ZvN-F2,RvN96}, see also Refs.~\cite{MV99,MM06}.
$F_\phi$ is the structure function for DIS by the exchange of a scalar $\phi$
coupling directly only to gluons via a term $\phi\,G^{\,\mu\nu\!}G_{\mu\nu}$ in
the Lagrangian (such as the Higgs boson in the limit of a five massless 
flavours and a very heavy top quark), where $G^{\,\mu\nu}$ denotes the gluon 
field strength tensor. 
The NNLO coefficient functions for the structure function $F_\phi$ have been 
calculated in Refs.~\cite{SMVV1,DGGL10}, while those for the corresponding 
fragmentation function $F_\phi^T$ are presented in Appendix A.

The spacelike physical kernels have been discussed, for the system $(\Ftwo,
F_\phi)$, in detail in Ref.~\cite{SMVV1}. The timelike case is completely
analogous up to a transposition of the matrices:
\beq
\label{eq:KTphit}
  {d \over d \ln \Qs} \; F^T \:\: = \:\:
  K^T \!\otimes F^T \:\: = \:\: 
  \Big\{ \!\Big( \beta\: \frac{d\:\! C^{\,T}}{d \as} 
  \:+\: C^{\,T}\!\otimes P^{\,T} \Big) \otimes (C^{\,T})^{\,-1} \Big)\! \Big\} 
  \otimes F^T
\eeq
 
\vspace*{-4mm}
\nin
with
 
\vspace*{-8mm}
\beq
\label{FKCmat}
  F^T \:=\: \left( \begin{array}{c} \!\FT\!\! \\[0.5mm] 
                                    \!F_\phi^T\!\! \end{array} \right)
  , \quad
  K^T \:=\: % \sum_{n=0} \as^{\,n+1} K^{T(n)} \;=\;
            \sum_{n=0} \as^{\,n+1} \! \left( \begin{array}{cc}
            \!  K_{\rm TT}^{(n)} & \! K_{\rm T\phi}^{(n)} \!\! \\[1mm]
            \!  K_{\rm \phi T}^{(n)} & \! K_{\rm \phi\phi}^{(n)T} \!\! 
            \end{array} \right)
  , \quad
  C^{\,T} \:=\: \sum_{n=0} \as^{\,n}
          \left( \begin{array}{cc}
            \!  C_{\,T,\rm q}^{(n)} \! & C_{\,\phi,\rm q}^{\,(n)T} \! \\[1mm]
            \!  C_{\,T,\rm g}^{(n)} \! & C_{\,\phi,\rm g}^{\,(n)T} \!
           \end{array} \right)
\eeq
where $\,c_{\,T,\rm q}^{(0)} \,=\, C_{\,\phi,\rm g}^{\,(0)T} \,=\, \delta\x1\,$
and $\,c_{\,T,\rm g}^{(0)} \,=\, C_{\,\phi,\rm q}^{\,(0)T} \,=\, 0$.  We have
skipped the superscript `{\it T'} where it is not needed for uniqueness in the 
present context. It should be noted that the normalization of 
$\,c_{\,T,\rm g}^{(n)}$ in Eq.~(\ref{FKCmat}) differs by a factor of 1/2 from
that in Refs.~\cite{RvN96,MM06}.
Finally $\beta$ in Eq.~(\ref{eq:KTphit}) is the standard beta 
function of QCD, $\beta \,=\, -\b0\, \as^{\,2} + \:\ldots\:$ 
with $\,\b0 = 11/3\:C_A - 2/3\: \nf$ and $\ca = N_{\rm colours} = 3$.

For the off-diagonal entries the analytic continuation involves, besides 
$x \ra 1/x$ and the multiplication by a factor $x$ due to the phase space of 
the detected parton in the SIA case \cite{RvN96}, a sign factor and a ratio of 
colour factors, leading to (with $C_F = 4/3$ in QCD)
\beq
\label{eq:ACphys2}
  \mbox{\it AC}\: \big[ K_{2\phi}^{(n)}(x) \big] \;=\; 
  - \,\frac{C_F}{\nf}\: x\, K_{2\phi}^{(n)}(1/x)
  \;\; , \quad
  \mbox{\it AC}\: \big[ K_{\phi 2}^{(n)}(x) \big] \;=\;
  - \,\frac{\nf}{C_F}\: x\, K_{\phi 2}^{(n)}(1/x) \; .
\eeq
The critical part of the analytic continuation is that of powers of $\,\ln\x1$,
which is given by
\beq
\label{eq:L1xac}
  \ln (1-x) \;\;\stackrel{AC_{\kappa}}{\longrightarrow}\;\; 
  \ln (1-x) \,-\, \ln x \,+\, \kappa\: i\:\!\pi 
  \quad \mbox{ with } \quad \kappa \,=\, 0 \:\mbox{ or }\: 1 \; .
\eeq
For $\kappa = 1$ the real part is taken in the end. It is not clear at all that
beyond NLO Eq.~(\ref{eq:L1xac}) is applicable, in either form, to quantities 
such as physical evolution kernels instead of to (classes of) Feynman diagrams,
see the discussions in Refs.~\cite{CFP80,FKL81,SV96}.

The NLO physical kernels for $(\Fone, F_\phi)$ and $(\FT, F_\phi^T)$ fulfil 
Eq.~(\ref{eq:ACphys1}) for both $\kappa = 0$ and $\kappa = 1$ in 
Eq.~(\ref{eq:L1xac}). However, for the NNLO diagonal entries we find
(restricting ourselves to $x < 1$)
\beq
\label{Pdiag2AC}
  {AC}_0^{} \: \big[ K_{11}^{\,(2)}(x) \big] \,-\, K_{TT}^{\,(2)}(x)
  \:\: = \:\: 24\,\z2\,\b0\,\cfs\; {1 + x^{\,2} \over 1-x}\; \ln x
  \:\: = \:\: 12\,\z2\,\b0\,\cf\, P_{\rm qq}^{\,(0)}(x) \, \ln x \:\: ,
\eeq
and a completely analogous relation with $\,P_{\rm gg}^{\,(0)} = 
K_{\phi\phi}^{\,(0)}\,$ on the right-hand-side for $\,K_{\phi\phi}^{\,(2)}$ and
$\,K_{\phi\phi}^{\,(2)T}$.
Using $\kappa = 1$ instead leads, besides $\z2\,\b0$ terms with 
$P_{\rm qq}^{\,(0)}(x)$ and $P_{\rm qq}^{\,(0)}(x) \ln x$,
to the same spurious $\z2\,\cft$ contribution as found in the analytic 
continuation in Ref.~\cite{MMV06}.
 
There is no obvious reason why an NNLO imperfection of the {\it AC} relation
should lead to an offset proportional to $\b0$ and the lowest-order kernel, 
and why exactly the same relation should hold for two different kernels. 
Hence it seems very likely that a non-vanishing r.h.s.~of Eq.~(\ref{Pdiag2AC})
is genuine and analogous to the `Crewther discrepancy' between the 
Gross--Llewellyn Smith sum rule in DIS and the Adler function in $e^+ e^-$ 
annihilation observed and discussed in Ref.~\cite{BK93}, see also the all-order
proof and the recent explicit fourth-order calculation in 
Refs.~\cite{Crewther}.
 
Therefore, if one tried to fix the so far unknown $\z2$ terms of the 
off-diagonal splitting functions by imposing Eq.~(\ref{eq:ACphys1}) for 
$\,K_{1\phi}^{\,(2)}$, $K_{\phi 1}^{\,(2)}$ and their timelike
counterparts, one should find an offset proportional to $\b0$ in the known
second moments which is then compensated by right-hand-sides analogous to 
that of Eq.~(\ref{Pdiag2AC}). Carrying out these calculations indeed leads to
\bea
\label{Pgq2AC2}
 \Big[ \, {AC}_0^{}\: \big[ K_{1\phi}^{\,(2)} \big] - K_{T\phi}^{\,(2)}
 \Big]_{N=2}
   &\!=\!&
 \z2\,\b0\,\cf \left(\, -\,\frak{212}{3}\,\ca \,-\, 16\,\cf \right) \; ,
\\[1mm]
\label{Pqg2AC2}
 \Big[ \, {AC}_0^{}\: \big[ K_{\phi 1}^{\,(2)} \big] - K_{\phi T}^{\,(2)}
 \Big]_{N=2}
   &\!=\!&
 \z2\,\b0\,\nf \left( \, 9\,\ca \,-\, \frak{38}{3}\,\cf + 4\,\b0 \right) \; .
\eea
For the required generalization of these relations to all values of $N$, we
consider also the $\kappa=1$ continuation which, while again leading to 
spurious non-$\b0$ terms, appears to provide the right correction terms for
Eq.~(\ref{Pgq2AC2}),
\beq
\label{Pgq2AC}
 {AC}_0^{}\: \big[ K_{1\phi}^{\,(2)} \big] - K_{T\phi}^{\,(2)} \:\:=\:\: 
  -\,6\,\z2\,\b0\, P_{\rm gq}^{\,(0)}(x) \left[ 2\,\ca ( 1- \ln x )
      \,+\, \cf \right] \; , 
\eeq
as well as the $\bs0$ part of Eq.~(\ref{Pqg2AC2}). 
We assume that also the first two terms on the r.h.s.~of Eq.~(\ref{Pqg2AC2}) 
correspond a to combination of $P_{\rm qg}^{\,(0)}$ and $P_{\rm qg}^{\,(0)} \ln x$.
(Poly-)$\:\!$logarithms with higher weight cannot occur since this is an
$\nf$ contribution and thus restricted to an overall weight of three. Imposing 
also other constraints discussed below we arrive at
\beq
\label{Pqg2AC}
 {AC}_0^{}\: \big[ K_{\phi 1}^{\,(2)} \big] - K_{\phi T}^{\,(2)} \:\:=\:\:
%    \z2\,\b0\, P_{\rm qg}^{\,(0)}(x) \left[ \,\ca ( 8 - 12 \ln x ) 
%
    \z2\,\b0\, P_{\rm qg}^{\,(0)}(x) \left[ 8\,(\ca - \cf) - 12 \ln x \:
      (\ca - 2\,\cf) \,+\, 6\,\b0\, \right] \; .
\eeq
Also the remaining uncertainty of the coefficients of $\ln x$ in this relation 
will be addressed below.

We are now ready to present the (only marginally provisional) results for the 
NNLO timelike splitting functions. 
For completeness we first recall the corresponding LO and NLO results:
\begin{eqnarray}
  \label{eq:PTqg0}
  P^{\,(0)T}_{\rm qg}(x) &\! = \!& P^{\,(0)S}_{\rm qg}(x) \:\: = \:\:
  \,2\,\*\colour4colour{\nf}\, \pqg(x) \:\: = \:\;
%%START
%%L %%texPTqg0 =
  2\,\*\colour4colour{\nf}\, \* ( 1 - 2\,\*x + 2\,\*x^{\,2} )
%%;
%%STOP
\; , 
\\[0.5mm]
  \label{eq:PTgq0}
  P^{\,(0)T}_{\rm gq}(x) &\! = \!& P^{\,(0)S}_{\rm gq}(x) \:\: = \:\:
  2\,\*\colour4colour{\cf}\, \pgq(x) \:\: = \:\,
%%START
%%L %%texPTgq0 =
  2\,\*\colour4colour{\cf} \* ( 2\,\* x^{-1} - 2 + x )
%%;
%%STOP
\end{eqnarray}
and
% 
%\pagebreak
%\nin
%
\begin{eqnarray}
  \label{eq:PTqg1} 
  \lefteqn{P^{\,(1)T}_{\rm qg}(x) \:\: = \:\:  
%%START
%%L %%texPTqg1 =
  4\,\*\colour4colour{\cf\*\nf}\,  \*  \big(
       - 6 + 23/2\,\*x - 10\,\*x^2
       - ( 5/2 - 2\,\*x - 2\,\*x^2)\, \*\H(0)
       - ( 1 - 2\,\*x + 4\,\*x^2)\, \*\Hh(0,0)
       + 2\, \*\H(1)
%%STOP
}
%%START 
  %%
  %%
  \nn\\[-0.5mm]&& \mbox{}
       + 2\*\pqg(x) \* ( 
         - \z2 
         + \H(1) / 2
         - 3\, \*\Hh(1,0)
         - \Hh(1,1)
         + \H(2)
         )
         \big)
  \,-\, 4/3\,\*\colour4colour{\nfs}\,  \*  \big(
         2
       + 2\,\* \pqg(x) \* (
           2/3
         + \H(0)
         - \H(1)
         )
         \big)
  \nn\\&& \mbox{}
  +4\,\*\colour4colour{\ca\*\nf}\,  \*  \big(
         - (20\,\*x^{-1} - 13 + 95\,\*x - 178\,\*x^2)/9
         - 4\,\*\z2\,\*x
         - ( 4 + 34\,\*x + 4\,\* x^2)/3\, \*\H(0) \,
  \nn\\&& \mbox{}
         + (2 + 12\,\*x)\, \*\Hh(0,0)
         - 2\,\* \H(1)
       - 2\*\pqg(-x) \* \Hh(-1,0)
       + 2\*\pqg(x) \* (
           2\, \* \Hh(1,0)
         - 5/6\,\* \H(1)
         + \Hh(1,1)
         - 2\,\* \H(2)
         )
         \big)
%%;
%%STOP
\:\: , \quad
%;
%%STOP
\\[2mm]
  \label{eq:PTgq1}
  \lefteqn{P^{\,(1)T}_{\rm gq}(x) \:\: = \:\:
%%START
%%L %%texPTgq1 =
  4\,\*\colour4colour{\cfs}\,  \*  \big(
         (9\,\*x - 1)/2
         - (8-x/2)\, \*\H(0)
         + (2-x)\,\* \Hh(0,0)
         - 2\,\*x\,\*\H(1)
         + 2\,\* \pgq(x)\,\* (
             2\,\* \Hh(1,0) 
           + \Hh(1,1) 
%%STOP
}
%%START
  %%
  %%
  \nn\\&& \mbox{}
           - 2\,\* \H(2)
           )
         \big)
  \,+\, 4\,\*\colour4colour{\cf\*\ca}\,  \*  \big(
         + 17/9\,\*x^{-1} + 5 - x - 44/9\,\*x^2 
         + 4\,\*\z2
         - ( 6\,\*x^{-1} - 8 - 9\,\*x - 8/3\,\*x^2 )\, \*\H(0)
  \nn\\&& \mbox{}
         - ( 8\,\*x^{-1} + 4 + 6\,\* x )\,\* \Hh(0,0)
         + 2\,\* x\,\* \H(1)
         - 2\,\*\pgq(-x)\,\* \Hh(-1,0)
         - 2\,\*\pgq(x)\,\* (
             3\,\* \Hh(1,0)
           + \Hh(1,1)
           - \H(2)
           ) 
         \big)
%%;
%%STOP
\;\; .
\end{eqnarray}
$H_{\vec{m}_1^{}}$ are the harmonic polylogarithms (HPLs) as defined in
Ref.~\cite{HPLs}
with $H_{0,1,0,1}(x) \equiv H_{2,2}$ etc.
Our new third-order contributions to Eq.~(\ref{eq:PTexp}) are given by  
\begin{eqnarray}
  \label{eq:PTqg2} 
  \lefteqn{P^{\,(2)T}_{\rm qg}(x) \;\; = \;\; } 
%%START
%%L %%texPTqg2 =
  \nn\\&& \mbox{}
\colour4colour{\cfs\*\nf}  \*  \big(
       - 227/4 - 1967/2\*x + 1069\*x^2
       - (   288 - 352\*x + 160\*x^2) \* \Hhh(1,0,0)
       + (34 - 32\*x - 4\*x^2) \* \Hh(1,1)
  \nn\\[-0.1mm]&& \mbox{}
       - ( 240 - 64\*x + 256\*x^2) \* \Hh(-2,0)
       - ( 180 - 192\*x + 144\*x^2) \* \Hhh(1,1,0)
       - ( 176 - 288\*x + 416\*x^2) \* \Hhh(2,0,0)
  \nn\\[-0.1mm]&& \mbox{}
       - ( 124 - 64\*x + 48\*x^2) \* \Hh(1,2)
       - ( 120 - 240\*x + 288\*x^2) \* \Hhh(2,1,0)
       - ( 104 - 208\*x + 224\*x^2) \* \Hh(2,2)
  \nn\\[-0.1mm]&& \mbox{}
       - ( 96 - 64\*x + 256\*x^2) \* \Hh(-3,0)
       - ( 76 - 64\*x + 48\*x^2) \* \Hhh(1,1,1)
       - ( 56 - 112\*x + 128\*x^2) \* \Hhh(2,1,1)
  \nn\\[-0.1mm]&& \mbox{}
       - ( 40 - 144\*x + 160\*x^2) \* \Hh(0,0)\*\z2
       - ( 28 + 144\*x - 16\*x^2) \* \H(3)
       - ( 2 - 116\*x + 324\*x^2) \* \H(2)
  \nn\\[-0.1mm]&& \mbox{}
       + (156 - 816\*x - 640\*x^2) \* \z3
       + (12 + 272\*x - 176\*x^2) \* \H(0)\*\z2
       + (16 - 256\*x + 128\*x^2) \* \Hhhh(0,0,0,0)
  \nn\\[-0.1mm]&& \mbox{}
       + (32 + 128\*x + 256\*x^2) \* \Hhh(-2,0,0)
       + (36 - 584\*x - 192\*x^2) \* \Hhh(0,0,0)
       + (64 + 320\*x + 256\*x^2) \* \Hhh(-1,0,0)
  \nn\\[-0.1mm]&& \mbox{}
       + (40 - 304\*x + 160\*x^2) \* \H(0)\*\z3
       + (56 - 1904/5\*x + 352/5\*x^2) \* \zs2
       + (64 - 128\*x + 192\*x^2) \* \Hh(3,1)
  \nn\\[-0.1mm]&& \mbox{}
       + (64 + 64\*x + 128\*x^2) \* (\H(-2)\*\z2 + 2\*\Hhh(-2,-1,0))
       - ( 496 + 240\*x - 224\*x^2) \* \Hh(-1,0)
  \nn\\[-0.1mm]&& \mbox{}
       + (92 - 1143\*x + 110\*x^2) \* \H(0)
       + (96 - 192\*x + 352\*x^2) \* \Hh(3,0)
       + (104 - 144\*x + 224\*x^2) \* \H(2)\*\z2
  \nn\\[-0.1mm]&& \mbox{}
       + (104 - 80\*x + 16\*x^2) \* \Hh(2,1)
       + (106 - 564\*x + 564\*x^2) \* \z2
       + (107 - 42\*x - 76\*x^2) \* \Hh(0,0)
  \nn\\[-0.1mm]&& \mbox{}
       + (108 - 64\*x - 80\*x^2) \* \H(1)\*\z2
       + (128 + 160\*x + 32\*x^2) \* (\H(-1)\*\z2 + 2\*\Hhh(-1,-1,0))
  \nn\\[-0.1mm]&& \mbox{}
       + (180 - 480\*x + 318\*x^2) \* \H(1)
       + (184 - 128\*x - 112\*x^2) \* \Hh(2,0)
       + (294 - 560\*x + 468\*x^2) \* \Hh(1,0)
  \nn\\[-0.1mm]&& \mbox{}
       + (8 - 16\*x + 32\*x^2) \* \H(4)
       - 96\* \pqg( - x) \* (\Hhh(-1,-2,0) + 2\*\Hhh(-1,-1,0,0) + \Hh(-1,0)\*\z2 - 11/6\*\Hhhh(-1,0,0,0) ) 
  \nn\\[-0.1mm]&& \mbox{}
       + 16\* \pqg(x) \* (
       31\*\H(1)\*\z3 
       - 6\*\Hhh(1,-2,0) 
       - 7\*\Hh(1,0)\*\z2 
       + 21\*\Hhh(1,0,0,0) 
       - \Hh(1,1)\*\z2 
       + \Hhhh(1,1,0,0) 
       + 3\*\Hhhh(1,1,1,0) 
  \nn\\[-0.1mm]&& \mbox{}
       + 2\*\Hhhh(1,1,1,1) 
       + 5\*\Hhh(1,1,2) 
       + 19\*\Hhh(1,2,0) 
       + 4\*\Hhh(1,2,1) 
       + 5\*\Hh(1,3)
          )
          \big)
  \nn\\[1mm]&& \mbox{\hspn}
+\colour4colour{\ca\*\cf\*\nf}  \*  \big(
       17597/36 - 220/3\*x^{-1} + 2659/6\*x - 3092/3\*x^2
       - ( 288 + 448\*x + 64\*x^2) \* \Hhh(-1,-1,0)
  \nn\\[-0.1mm]&& \mbox{}
       - ( 592/3 - 1352/3\*x + 192\*x^2) \* \Hh(2,1)
       - ( 2422/27 - 30203/27\*x + 17918/27\*x^2) \* \H(0)
  \nn\\[-0.1mm]&& \mbox{}
       - ( 188 - 32\*x^{-1} - 272\*x + 120\*x^2) \* \H(1)\*\z2
       - ( 1007/9 + 6254/9\*x - 2444/9\*x^2) \* \Hh(0,0)
  \nn\\[-0.1mm]&& \mbox{}
       - ( 160 + 160\*x + 320\*x^2) \* \Hh(3,0)
       - ( 176 - 192\*x + 384\*x^2) \* \Hh(3,1)
       - ( 96 + 160\*x - 32\*x^2) \* \H(-1)\*\z2
  \nn\\[-0.1mm]&& \mbox{}
       - ( 64 + 32\*x + 128\*x^2) \* (\H(-2)\*\z2 + 2\*\Hhh(-2,-1,0))
       + (28/3 + 2032/3\*x + 280/3\*x^2) \* \Hh(2,0)
  \nn\\[-0.1mm]&& \mbox{}
       - ( 488/9 - 304/9\*x^{-1} - 868/9\*x + 1904/9\*x^2) \* \Hh(1,1)
       - ( 148/3 - 1832/3\*x - 640/3\*x^2) \* \Hhh(0,0,0)
  \nn\\[-0.1mm]&& \mbox{}
       - ( 48 + 64\*x + 64\*x^2) \* \Hh(-1,2)
       - ( 32 + 64\*x) \* \Hhhh(0,0,0,0)
       - ( 16 + 96\*x^{-1} + 504\*x - 368\*x^2) \* \Hhh(1,0,0)
  \nn\\[-0.1mm]&& \mbox{}
       - ( 44/5 - 2632/5\*x + 128\*x^2) \* \zs2
       - ( 56 - 16\*x + 160\*x^2) \* \H(0)\*\z3
       + (16 - 80\*x - 144\*x^2) \* \Hhh(-1,0,0)
  \nn\\[-0.1mm]&& \mbox{}
       + (32 - 256\*x + 256\*x^2) \* \Hh(0,0)\*\z2
       + (64 + 96\*x + 128\*x^2) \* \H(4)
       + (28/9 + 4540/9\*x - 2620/3\*x^2) \* \z2
  \nn\\[-0.1mm]&& \mbox{}
       + (268/3 - 32\*x^{-1} - 608/3\*x + 728/3\*x^2) \* \Hh(1,2)
       + (-260/3 - 176/3\*x - 8\*x^2) \* \H(0)\*\z2
  \nn\\[-0.1mm]&& \mbox{}
       + (2788/27 - 80/27\*x^{-1} + 19660/27\*x - 19970/27\*x^2) \* \H(1)
       + (388/3 + 256/3\*x - 40/3\*x^2) \* \H(3)
  \nn\\[-0.1mm]&& \mbox{}
       + (136 - 1040\*x + 672\*x^2) \* \Hhh(2,0,0)
       + (136 - 368\*x + 320\*x^2) \* \Hhh(2,1,1)
       - ( 200 - 592\*x + 512\*x^2) \* \H(2)\*\z2
  \nn\\[-0.1mm]&& \mbox{}
       + (424/3 - 32/3\*x^{-1} - 536/3\*x + 544/3\*x^2) \* \Hhh(1,1,1)
       + (2204/9 - 3772/9\*x + 3176/3\*x^2) \* \H(2)
  \nn\\[-0.1mm]&& \mbox{}
       + (452/3 - 32\*x^{-1} - 1024/3\*x + 1048/3\*x^2) \* \Hhh(1,1,0)
       + (168 - 624\*x + 448\*x^2) \* \Hh(2,2)
  \nn\\[-0.1mm]&& \mbox{}
       + (184 - 656\*x + 512\*x^2) \* \Hhh(2,1,0)
       + (144 + 32\*x + 384\*x^2) \* \Hh(-3,0)
       + (336 + 16\*x + 160\*x^2) \* \Hh(-2,0)
  \nn\\[-0.1mm]&& \mbox{}
       + (1340/3 + 64\*x^{-1} + 4904/3\*x + 488\*x^2) \* \z3
       - ( 212 - 304/3\*x^{-1} - 2036/3\*x + 3284/3\*x^2) \* \Hh(1,0)
  \nn\\[-0.1mm]&& \mbox{}
       + (672 + 544\*x - 16\*x^2) \* \Hh(-1,0)
       - ( 96\*x + 128\*x^2) \* \Hhh(-2,0,0)
       - 8\*\pqg( - x) \* (
       10\*\Hhh(-1,-2,0) 
       - 8\*\Hhhh(-1,-1,0,0) 
  \nn\\[-0.1mm]&& \mbox{}
       - 10\*\Hh(-1,-1)\*\z2 
       + 5\*\H(-1)\*\z3 
       - 12\*\Hhhh(-1,-1,-1,0) 
       + 4\*\Hhh(-1,-1,2)
       + 3\*\Hhhh(-1,0,0,0) 
       - 12\*\Hhh(-1,2,0) 
       - 4\*\Hhh(-1,2,1) 
       )
  \nn\\[-0.1mm]&& \mbox{} 
       - 8\*\pqg(x) \* (  
       99\*\H(1)\*\z3 
       - 10\*\Hhh(1,-2,0) 
       - 10\*\Hh(1,0)\*\z2 
       + 7\*\Hhhh(1,0,0,0)
       - 4\*\Hh(1,1)\*\z2 
       - 26\*\Hhhh(1,1,0,0) 
       + 2\*\Hhhh(1,1,1,0) 
  \nn\\[-0.1mm]&& \mbox{}  
       + 8\*\Hhhh(1,1,1,1) 
       + 6\*\Hhh(1,1,2) 
       + 30\*\Hhh(1,2,0) 
       + 4\*\Hhh(1,2,1) 
       - 14\*\Hh(1,3)
       )
          \big)
  \nn\\[1mm]&& \mbox{\hspn}
+\colour4colour{\cas\*\nf}  \*  \big(
       2057/9 + 2092/81\*x^{-1} - 672\*x + 53753/81\*x^2  
       - ( 196 + 2416/3\*x - 328/3\*x^2) \* \z3
  \nn\\[-0.1mm]&& \mbox{}  
       - ( 3940/27 - 752/27\*x^{-1} + 8416/27\*x - 7448/27\*x^2) \* \H(1)
       + (288\*x - 96\*x^2) \* (\Hhh(2,1,0) + \Hh(2,2))
  \nn\\[-0.1mm]&& \mbox{}  
       - ( 1262/9 - 112/9\*x^{-1} - 4828/9\*x + 2900/3\*x^2) \* \H(2)
       - ( 100 - 64/3\*x^{-1} + 312\*x - 72\*x^2) \* \Hh(2,0)
  \nn\\[-0.1mm]&& \mbox{}  
       - ( 352/3 + 64/3\*x^{-1} + 704/3\*x + 80\*x^2) \* \Hh(-1,2)
       - ( 100 + 776\*x - 96\*x^2) \* \H(4)
       - ( 64 + 128\*x^2) \* \Hh(-2,2)
  \nn\\[-0.1mm]&& \mbox{}  
       - ( 2996/27 - 40\*x^{-1} + 4832/27\*x - 5044/9\*x^2) \* \H(0)
       - ( 104 - 432\*x + 32\*x^2) \* \Hhh(-2,0,0)
  \nn\\[-0.1mm]&& \mbox{}  
       - ( 80 - 256\*x + 192\*x^2) \* \Hhh(2,1,1)
       - ( 598/9 + 128/9\*x^{-1} + 3508/9\*x + 7208/9\*x^2) \* \Hh(0,0)
  \nn\\[-0.1mm]&& \mbox{}  
       - ( 196/3 - 32/3\*x^{-1} - 344/3\*x + 400/3\*x^2) \* \Hhh(1,1,1)
       + (168 - 432\*x + 384\*x^2) \* \H(2)\*\z2
  \nn\\[-0.1mm]&& \mbox{}  
       - ( 310/3 - 544/9\*x^{-1} - 308\*x + 536/3\*x^2) \* \z2
       - ( 56 - 496\*x) \* \Hh(-3,0)
       + (64 - 1264\*x) \* \Hhhh(0,0,0,0)
  \nn\\[-0.1mm]&& \mbox{}  
       - ( 152/3 + 32/3\*x^{-1} + 256/3\*x + 248\*x^2) \* \Hh(-1,0)
       - ( 28 - 632\*x + 96\*x^2) \* \Hh(0,0)\*\z2
  \nn\\[-0.1mm]&& \mbox{}  
       + (96 + 64/3\*x^{-1} - 484/3\*x + 824/3\*x^2) \* \H(0)\*\z2
       + (134/9 - 128/9\*x^{-1} - 700/9\*x + 652/3\*x^2) \* \Hh(1,1)
  \nn\\[-0.1mm]&& \mbox{}  
       + (16 - 352\*x - 64\*x^2) \* \Hhh(-2,-1,0)
       + (728/3 + 224/3\*x^{-1} + 188/3\*x - 304/3\*x^2) \* \Hhh(1,0,0)
  \nn\\[-0.1mm]&& \mbox{}  
       + (64/3 - 128/3\*x^{-1} + 2680/3\*x - 192\*x^2) \* \Hhh(0,0,0)
       + (64/3 - 64/3\*x^{-1} + 292\*x + 120\*x^2) \* \H(3)
  \nn\\[-0.1mm]&& \mbox{}  
       + (34 + 252/5\*x + 704/5\*x^2) \* \zs2
       + (160/3 + 32\*x^{-1} + 304/3\*x - 472/3\*x^2) \* (\Hhh(1,1,0) + \Hh(1,2))
  \nn\\[-0.1mm]&& \mbox{}  
       + (184/3 + 64/3\*x^{-1} - 344/3\*x + 832/3\*x^2) \* \Hh(-2,0)
       + (64 - 96\*x + 160\*x^2) \* \Hh(3,0)
  \nn\\[-0.1mm]&& \mbox{}  
       + (440/3 - 64/3\*x^{-1} - 664/3\*x + 464/3\*x^2) \* \H(1)\*\z2
       + (72 - 176\*x + 96\*x^2) \* \H(-2)\*\z2
  \nn\\[-0.1mm]&& \mbox{}  
       + (92 + 408\*x - 32\*x^2) \* \Hhh(2,0,0)
       + (320/3 - 64/3\*x^{-1} + 112/3\*x - 208\*x^2) \* \Hhh(-1,-1,0)
  \nn\\[-0.1mm]&& \mbox{}  
       + (128 + 32/3\*x^{-1} - 400\*x + 640/3\*x^2) \* \Hh(2,1)
       + (16 + 960\*x) \* \H(0)\*\z3
       + (176 - 192\*x + 320\*x^2) \* \Hh(3,1)
  \nn\\[-0.1mm]&& \mbox{}  
       + (512/3 + 32/3\*x^{-1} + 760/3\*x - 24\*x^2) \* \H(-1)\*\z2
       - ( 562/9 + 8/9\*x^{-1} - 92/9\*x - 704/3\*x^2) \* \Hh(1,0)
  \nn\\[-0.1mm]&& \mbox{}  
       - ( 872/3 - 128/3\*x^{-1} + 520/3\*x - 304/3\*x^2) \* \Hhh(-1,0,0)
       + 8\*\pqg( - x) \* ( 
       17\*\H(-1)\*\z3 
       + 14\*\Hhh(-1,-2,0) 
  \nn\\[-0.1mm]&& \mbox{}
       - 26\*\Hh(-1,-1)\*\z2 
       - 12\*\Hhhh(-1,-1,-1,0) 
       + 8\*\Hhhh(-1,-1,0,0)
       + 20\*\Hhh(-1,-1,2) 
       + 14\*\Hh(-1,0)\*\z2 
       + \Hhhh(-1,0,0,0)
  \nn\\[-0.1mm]&& \mbox{}
       - 8\*\Hhh(-1,2,0) 
       - 4\*\Hhh(-1,2,1) 
       )
       + 8\*\pqg(x) \* ( 
       31\*\H(1)\*\z3 
       - 2\*\Hhh(1,-2,0) 
       + 6\*\Hh(1,0)\*\z2 
       - 15\*\Hhhh(1,0,0,0) 
       - 2\*\Hh(1,1)\*\z2 
  \nn\\[-0.1mm]&& \mbox{} 
       - 20\*\Hhhh(1,1,0,0)
       - 4\*\Hhhh(1,1,1,0) 
       + 4\*\Hhhh(1,1,1,1) 
       - 4\*\Hhh(1,1,2) 
       - 4\*\Hhh(1,2,1) 
       - 16\*\Hh(1,3)
       )
          \big)
  \nn\\[1mm]&& \mbox{\hspn}
+\colour4colour{\cf\*\nfs}  \*  \big(
       - 4847/54 + 200/27\*x^{-1} - 2375/27\*x + 4066/27\*x^2
       - ( 32 - 64\*x) \* \Hhhh(0,0,0,0)
  \nn\\[-0.1mm]&& \mbox{}
       - ( 416/9 - 376/9\*x + 88/3\*x^2) \* \H(2)
       + (1684/27 + 512/27\*x^{-1} - 4022/27\*x - 4480/27\*x^2) \, \* \H(0)
  \nn\\[-0.1mm]&& \mbox{}
       - ( 808/27 - 560/27\*x + 248/27\*x^2) \* \H(1)
       + (902/9 + 128/9\*x^{-1} + 956/9\*x + 1400/9\*x^2) \* \Hh(0,0)
  \nn\\[-0.1mm]&& \mbox{}
       - ( 64/3 - 128/3\*x + 176/3\*x^2) \* \H(3)
       + (272/9 - 64/9\*x^{-1} + 8/9\*x + 32\*x^2) \* \z2
  \nn\\[-0.1mm]&& \mbox{}
       + (48 - 64/9\*x^{-1} - 496/9\*x^2) \* \Hh(-1,0)
       + (344/9 - 376/9\*x + 328/9\*x^2) \* \Hh(1,1)
  \nn\\[-0.1mm]&& \mbox{}
       + (80 - 160\*x + 176\*x^2)/3 \, \* \Hh(2,0)
       - (32 - 112\*x - 96\*x^2)/3 \, \* \Hhh(0,0,0)
       + (64 - 128\*x + 144\*x^2)/3 \,  \* \Hh(2,1)
  \nn\\[-0.1mm]&& \mbox{}
       + (32 - 64\*x + 64/3\*x^2) \* \Hh(-2,0)
       + ( 80 - 352\*x + 144\*x^2)/3 \, \* \H(0)\*\z2
       + (304/3 - 608/3\*x + 160\*x^2) \* \z3
  \nn\\[-0.1mm]&& \mbox{}
       + (192 - 248\*x + 232\*x^2)/3 \,  \* \Hh(1,0)
       - 8/3\*\pqg(x) \* (
       3\*\H(1)\*\z2 
       + 5\*\Hh(1,2)
       + 7\*\Hhh(1,1,0)
       + 5\*\Hhh(1,1,1) 
       - 12\*\Hhh(1,0,0) 
       )
          \big)
  \nn\\[1mm]&& \mbox{\hspn}
+\colour4colour{\ca\*\nfs}  \*  \big(
       - 14/9 - 44/9\*x^{-1} - 1216/9\*x + 916/9\*x^2
       - ( 40 - 352/3\*x + 272/3\*x^2) \* \z3
  \nn\\[-0.1mm]&& \mbox{}
       - ( 296/9 + 64/9\*x^{-1} + 416/9\*x + 536/9\*x^2) \* \Hh(0,0)
       - ( 32 - 96\*x + 224/3\*x^2) \* \Hh(2,1)
  \nn\\[-0.1mm]&& \mbox{}
       - ( 64/3 + 640/3\*x) \* \Hhh(0,0,0)
       + (44/27 - 256/27\*x^{-1} + 752/27\*x - 2212/27\*x^2) \* \H(0)
  \nn\\[-0.1mm]&& \mbox{}
       + (100/3 + 32/9\*x^{-1} + 208/3\*x - 16/9\*x^2) \* \z2
       - ( 24 - 112/3\*x + 112/3\*x^2) \* \H(0)\*\z2
  \nn\\[-0.1mm]&& \mbox{}
       - ( 320/9 - 32/9\*x^{-1} - 544/9\*x + 200/3\*x^2) \* \Hh(1,1)
       - ( 64/3 - 128/3\*x + 64/3\*x^2) \* \Hh(-2,0)
  \nn\\[-0.1mm]&& \mbox{}
       + (16/9 + 32/3\*x^{-1} + 496/9\*x - 664/9\*x^2) \* \Hh(1,0)
       + (356/9 - 1360/9\*x + 632/9\*x^2) \* \H(2)
  \nn\\[-0.1mm]&& \mbox{}
       + (592 - 320\*x^{-1} - 1232\*x + 2632\*x^2)/27 \, \* \H(1)
       + (104/3 - 16\*x + 176/3\*x^2) \* \H(3)
  \nn\\[-0.1mm]&& \mbox{}
       - ( 144 - 32\*x^{-1} - 96\*x - 512\*x^2)/9 \, \* \Hh(-1,0)
       + (8 + 80\*x - 16\*x^2) \* \Hh(2,0)
       + 32/3\*\pqg( - x) \* (
       2\*\H(-1)\*\z2 
  \nn\\[-0.1mm]&& \mbox{}
       + 2\*\H(-1,-1,0) 
       + \Hhh(-1,0,0) 
       - \Hh(-1,2)
       )
       - 8/3\*\pqg(x) \* (
       4\*\H(1)\*\z2 
       + 19\*\Hhh(1,0,0) 
       + 2\*\Hhh(1,1,0) 
       - 5\*\Hhh(1,1,1) 
       + 2\*\Hh(1,2) 
       )
          \big)
  \nn\\[1mm]&& \mbox{\hspn}
+\colour4colour{\nft}  \*  \big(
%*       - 8/9 + 16/3\*x - 16/3\*x^2 
%*       + (40/9 - 32/9\*x + 32/9\*x^2) \* (\H(0) - \H(1))
       16/9 + 8/3 \* [ \H(0) - \H(1) 
%  %%
%  %%
%  \nn\\[-0.1mm]&& \mbox{}
       + \pqg(x) \* (
         \z2 - 1 
         + 2/3 \* (\H(0) - \H(1))
       + \Hh(0,0) 
       - \Hh(1,0) 
       + \Hh(1,1) 
       - \H(2) 
       ) ] 
          \big)
%%;
%%STOP
\end{eqnarray}
% 
%\vspace*{-2mm}
%\nin
and
% 
%\vspace*{-6mm}
\begin{eqnarray}
  \label{eq:PTgq2} 
  \lefteqn{P^{\,(2)T}_{\rm gq}(x) \;\; = \;\; } 
%%START
%%L %%texPTgq2 =
  \nn\\&& \mbox{}
\colour4colour{\cft}  \*  \big(
       - 1915/2 + 794\*x^{-1} + 731/4\*x
       - ( 464 - 320\*x^{-1} - 200\*x) \* \H(2)\*\z2
       - ( 290 + 251\*x) \* \Hh(0,0)
  \nn\\[-0.1mm]&& \mbox{}
       - ( 352 - 144\*x^{-1} - 252\*x) \* \H(1)\*\z2
       - ( 264 - 180\*x^{-1} - 140\*x) \* \Hh(1,0)
       - ( 240 - 256\*x^{-1} - 120\*x) \* \Hh(3,1)
  \nn\\[-0.1mm]&& \mbox{}
       - ( 230 - 228\*x^{-1} - 16\*x) \* \H(1)
       - ( 224 + 256\*x^{-1} - 148\*x) \* \z2
       - ( 176 - 128\*x^{-1} - 88\*x) \* \Hh(3,0)
  \nn\\[-0.1mm]&& \mbox{}
       - ( 776/5 - 1216/5\*x^{-1} - 84/5\*x) \* \zs2
       - ( 128 - 64\*x) \* \H(4)
       - ( 104 - 40\*x) \* \Hh(2,0)
       + (248 - 28\*x) \* \H(3)
  \nn\\[-0.1mm]&& \mbox{}
       - ( 96 + 720\*x^{-1} + 196\*x) \* \z3
       - ( 80 + 256\*x^{-1} - 208\*x) \* \Hh(-1,0)
       - ( 80 - 128\*x^{-1} + 200\*x) \* \H(0)\*\z3
  \nn\\[-0.1mm]&& \mbox{}
       - ( 72 - 60\*x^{-1} - 80\*x) \* \Hh(1,1)
       - ( 72 - 56\*x) \* \Hh(2,1)
       - ( 64 + 32\*x) \* (\H(-1)\*\z2 + 2\*\Hhh(-1,-1,0))
  \nn\\[-0.1mm]&& \mbox{}
       - ( 64 + 96\*x) \* \Hh(-3,0)
       - ( 64 + 64\*x) \* \Hh(0,0)\*\z2
       - ( 64 + 128\*x^{-1} + 64\*x) \* \Hhh(-2,0,0)
       + (208 - 140\*x) \* \H(2)
  \nn\\[-0.1mm]&& \mbox{}
       - ( 32 - 48\*x^{-1} - 4\*x) \* \Hhh(1,1,0)
       - ( 32 - 96\*x) \* \Hhhh(0,0,0,0)
       - ( 24 + 192\*x^{-1} + 20\*x) \* \H(0)\*\z2
  \nn\\[-0.1mm]&& \mbox{}
       + (197 + 208\*x^{-1} + 29\*x) \* \H(0)
       + (16 - 52\*x) \* \Hhh(1,1,1)
       + (128 + 128\*x^{-1} + 96\*x) \* (\H(-2)\*\z2 + 2\*\Hhh(-2,-1,0))
  \nn\\[-0.1mm]&& \mbox{}
       + (128 - 192\*x^{-1} - 112\*x) \* \Hh(-2,0)
       + (128 - 48\*x^{-1} - 60\*x) \* \Hh(1,2)
       + (128 + 48\*x^{-1} - 36\*x) \* \Hhh(1,0,0)
  \nn\\[-0.1mm]&& \mbox{}
       + (80 - 64\*x^{-1} - 40\*x) \* \Hhh(2,1,0)
       + (144 - 64\*x^{-1} - 72\*x) \* \Hh(2,2)
       + (160 + 40\*x) \* \Hhh(0,0,0)
  \nn\\[-0.1mm]&& \mbox{}
       - ( 16 - 128\*x^{-1} - 40\*x) \* \Hhh(2,0,0)
       + (208 - 192\*x^{-1} - 104\*x) \* \Hhh(2,1,1)
       + (224 + 96\*x^{-1} + 96\*x) \* \Hhh(-1,0,0)
  \nn\\[-0.1mm]&& \mbox{}
       + 16\* \pgq( - x) \* (2\*\Hhh(-1,-2,0) + 4\*\Hhhh(-1,-1,0,0) 
          - \Hhhh(-1,0,0,0))
       + 16\*\pgq(x) \*  (
       2\*\H(1)\*\z3 
       - 2\*\Hhh(1,-2,0) 
       + 4\*\Hh(1,0)\*\z2
  \nn\\[-0.1mm]&& \mbox{}
       - 6\*\Hhhh(1,0,0,0) 
       + 3\*\Hh(1,1)\*\z2 
       - 6\*\Hhhh(1,1,0,0) 
       + \Hhhh(1,1,1,0) 
       + 2\*\Hhhh(1,1,1,1) 
       - 3\*\Hhh(1,1,2) 
       + \Hhh(1,2,0) 
       - 4\*\Hhh(1,2,1) 
       - 8\*\Hh(1,3)
       )
          \big)
  \nn\\[1mm]&& \mbox{\hspn}
+\colour4colour{\ca\*\cfs}  \*  \big(
         1735/6 + 200\*x^{-1} - 4811/12\*x - 140/3\*x^2
       - ( 336 - 448\*x^{-1} - 312\*x) \* \Hh(2,2)
  \nn\\[-0.1mm]&& \mbox{}
       - ( 2338/3 - 80\*x^{-1} + 441\*x + 1408/9\*x^2) \* \Hh(0,0)
       - ( 1760/3 - 336\*x^{-1} - 832/3\*x + 64/3\*x^2) \* \Hhh(1,0,0)
  \nn\\[-0.1mm]&& \mbox{}
       - ( 336 - 384\*x^{-1} - 216\*x) \* \Hhh(2,1,0)
       - ( 8980/9 + 5800/9\*x^{-1} + 4576/9\*x + 352/3\*x^2) \* \H(2)
  \nn\\[-0.1mm]&& \mbox{}
       - ( 976/3 - 904/3\*x^{-1} - 404/3\*x + 32\*x^2) \* \Hh(1,2)
       - ( 320 - 256\*x^{-1} - 208\*x) \* \Hhh(2,0,0)
  \nn\\[-0.1mm]&& \mbox{}
       - ( 272 + 832\*x^{-1} + 376\*x) \* \H(0)\*\z3
       - ( 272 - 320\*x^{-1} - 184\*x) \* \Hhh(2,1,1)
       - ( 208 + 448\*x^{-1} + 344\*x) \* \Hh(3,0)
  \nn\\[-0.1mm]&& \mbox{}
       - ( 192 + 48\*x^{-1} + 72\*x) \* \Hhh(-1,0,0)
       - ( 144 - 96\*x^{-1} - 128\*x) \* \Hh(-2,0)
       - ( 128 - 32\*x^{-1} + 16\*x) \* \H(-2)\*\z2
  \nn\\[-0.1mm]&& \mbox{}
       - ( 128 + 64\*x^{-1} + 96\*x) \* \Hhh(-2,-1,0)
       - ( 256/3 - 88\*x^{-1} - 332/3\*x + 32/3\*x^2) \* \Hhh(1,1,0)
  \nn\\[-0.1mm]&& \mbox{}
       - ( 608/5 + 3664/5\*x^{-1} + 2144/5\*x) \* \zs2
       - ( 88 - 192\*x^{-1} + 268\*x) \* \Hh(-1,0)
       - ( 32 - 48\*x) \* \Hh(-3,0)
  \nn\\[-0.1mm]&& \mbox{}
       - ( 152/3 - 160/3\*x^{-1} - 232/3\*x + 32/3\*x^2) \* \Hhh(1,1,1)
       + (552 + 264\*x^{-1} + 424\*x + 128/3\*x^2) \* \z3
  \nn\\[-0.1mm]&& \mbox{}
       + ( 136/3 + 640/3\*x^{-1} - 680/3\*x - 32\*x^2) \* \H(3)
       - ( 160/9 - 848/9\*x^{-1} + 280/9\*x + 352/9\*x^2) \* \Hh(1,1)
  \nn\\[-0.1mm]&& \mbox{}
       - ( 64 + 24\*x^{-1} + 32\*x) \* \H(-1)\*\z2
       - ( 16 + 128\*x^{-1} + 40\*x) \* \Hh(0,0)\*\z2
       - ( 16 + 320\*x^{-1} + 216\*x) \* \Hh(3,1)
  \nn\\[-0.1mm]&& \mbox{}
       + (72 + 192\*x^{-1} + 128\*x - 32/3\*x^2) \* \H(0)\*\z2
       + (368/3 - 268/3\*x + 64/3\*x^2) \* \Hhh(0,0,0)
  \nn\\[-0.1mm]&& \mbox{}
       + (424/3 - 464/3\*x^{-1} + 28/3\*x + 32\*x^2) \* \Hh(2,1)
       + (600 - 1280/3\*x^{-1} + 96\*x + 224/3\*x^2) \* \Hh(2,0)
  \nn\\[-0.1mm]&& \mbox{}
       + (488/3 + 620/9\*x^{-1} - 200\*x - 704/9\*x^2) \* \Hh(1,0)
       + (192 + 48\*x^{-1} + 128\*x) \* \Hhh(-1,-1,0)
  \nn\\[-0.1mm]&& \mbox{}
       + (272 - 608\*x^{-1} - 456\*x) \* \H(2)\*\z2
       + (336 + 512\*x^{-1} + 376\*x) \* \H(4)
       + (160 + 48\*x^{-1} + 96\*x) \* \Hh(-1,2)
  \nn\\[-0.1mm]&& \mbox{}
       - ( 32 + 32\*x) \* (\Hhh(-2,0,0) - 2\*\Hhhh(0,0,0,0))
       + (40301/27 + 244\*x^{-1} + 18911/27\*x + 8932/27\*x^2) \* \H(0)
  \nn\\[-0.1mm]&& \mbox{}
       + (672 - 1760/3\*x^{-1} - 180\*x + 224/3\*x^2) \* \H(1)\*\z2
       + (3452/3 + 528\*x^{-1} + 436/3\*x + 1760/9\*x^2) \* \z2
  \nn\\[-0.1mm]&& \mbox{}
       + (26008/27 - 39092/27\*x^{-1} + 3988/27\*x + 5908/27\*x^2) \* \H(1)
       + 8\*\pgq( - x) \* (
       4\*\Hh(-1,3)
       + 11\*\H(-1)\*\z3 
  \nn\\[-0.1mm]&& \mbox{}
       - 18\*\Hh(-1,-1)\*\z2 
       - 8\*\Hhh(-1,2,0) 
       - 4\*\Hhh(-1,2,1) 
       - 4\*\Hhhh(-1,-1,0,0)
       + 12\*\Hhh(-1,-1,2) 
       + 12\*\Hh(-1,0)\*\z2 
       + \Hhhh(-1,0,0,0)
  \nn\\[-0.1mm]&& \mbox{}
       + 10\*\Hhh(-1,-2,0) 
       - 12\*\Hhhh(-1,-1,-1,0) 
          )
       - 8\*\pgq(x) \* (  
       4\*\Hh(-2,2) 
       + 51\*\H(1)\*\z3 
       - 10\*\Hhhh(1,1,0,0) 
       - 14\*\Hh(1,3) 
       - 4\*\Hhh(1,2,1) 
  \nn\\[-0.1mm]&& \mbox{} 
       + 3\*\Hhhh(1,0,0,0) 
       + 12\*\Hh(1,1)\*\z2
       - 6\*\Hhh(1,-2,0) 
       - 2\*\Hh(1,0)\*\z2
       + 8\*\Hhhh(1,1,1,1) 
       + 2\*\Hhh(1,1,2) 
       + 34\*\Hhh(1,2,0) 
       + 14\*\Hhhh(1,1,1,0) 
       )
          \big)
  \nn\\[1mm]&& \mbox{\hspn}
+\colour4colour{\cas\*\cf}  \*  \big(
       - ( 12 + 752/3\*x^{-1} - 532/3\*x + 128/3\*x^2) \* \H(0)\*\z2
       - ( 152 + 128\*x^{-1} + 92\*x) \* \H(4)
  \nn\\[-0.1mm]&& \mbox{}
       - ( 22270/27 - 31504/27\*x^{-1} + 688/27\*x + 1924/9\*x^2) \* \H(1)
       - ( 144 + 64\*x^{-1} + 8\*x) \* \H(-2)\*\z2
  \nn\\[-0.1mm]&& \mbox{}
       - ( 17798/27 - 3514/9\*x^{-1} + 3436/27\*x - 4136/9\*x^2)
       - ( 400 - 160\*x^{-1} - 148\*x + 640/3\*x^2) \* \z3
  \nn\\[-0.1mm]&& \mbox{}
       - ( 576 - 1504/3\*x^{-1} + 424\*x + 512/3\*x^2) \* \Hh(2,0)
       - ( 1292/3 + 32/3\*x^{-1} + 284/3\*x + 128/3\*x^2) \* \H(3)
  \nn\\[-0.1mm]&& \mbox{}
       - ( 848/3 - 1856/3\*x^{-1} + 3344/3\*x + 224/3\*x^2) \* \Hhh(0,0,0)
       + (572/5 + 1984/5\*x^{-1} + 314\*x) \* \zs2
  \nn\\[-0.1mm]&& \mbox{}
       - ( 32612/27 - 13384/27\*x^{-1} + 23630/27\*x + 12964/27\*x^2) \* \H(0) 
       - ( 160 - 48\*x) \* \Hhh(-2,-1,0)
  \nn\\[-0.1mm]&& \mbox{}
       - ( 168 + 176/3\*x^{-1} + 160\*x + 32/3\*x^2) \* \H(-1)\*\z2
       - ( 280 - 344\*x^{-1} + 32\*x + 64\*x^2) \* \H(1)\*\z2
  \nn\\[-0.1mm]&& \mbox{}
       - ( 848 + 5176/9\*x^{-1} + 2800/9\*x + 352/9\*x^2) \* \z2
       + (152 - 128\*x^{-1} - 156\*x) \* \Hh(0,0)\*\z2
  \nn\\[-0.1mm]&& \mbox{}
       - ( 1280/9 + 248/3\*x^{-1} - 1052/9\*x - 704/9\*x^2) \* \Hh(-1,0)
       - ( 128 - 512\*x^{-1} - 632\*x) \* \Hhhh(0,0,0,0)
  \nn\\[-0.1mm]&& \mbox{}
       - ( 328/3 + 272\*x^{-1} - 544/3\*x + 128/3\*x^2) \* \Hhh(-1,0,0)
       + (128 - 192\*x^{-1} - 112\*x) \* \Hhh(2,1,0)
  \nn\\[-0.1mm]&& \mbox{}
       + (384 + 576\*x^{-1} + 480\*x) \* (\Hh(3,0) + 1/3\*\Hh(3,1))
       - ( 296/3 - 448/3\*x^{-1} + 260/3\*x + 128/3\*x^2) \* \Hh(2,1)
  \nn\\[-0.1mm]&& \mbox{}
       - ( 280/3 + 752/3\*x^{-1} - 928/3\*x + 256/3\*x^2) \* \Hh(-2,0)
       + (64 - 128\*x^{-1} - 80\*x) \* \Hhh(2,1,1)
  \nn\\[-0.1mm]&& \mbox{}
       - ( 16 - 352/3\*x^{-1} + 160\*x - 64/3\*x^2) \* \Hhh(-1,-1,0)
       + (1300/3 - 1024/3\*x^{-1} - 1004/3\*x) \* \Hhh(1,0,0)
  \nn\\[-0.1mm]&& \mbox{}
       + (104/3 - 160/3\*x^{-1} - 76/3\*x + 32/3\*x^2) \* \Hhh(1,1,1)
       + (48 + 320\*x^{-1} + 248\*x) \* \H(2)\*\z2
  \nn\\[-0.1mm]&& \mbox{}
       + (11848/9 + 3920/9\*x^{-1} + 7280/9\*x + 3688/9\*x^2) \* \Hh(0,0)
       + (112 - 256\*x^{-1} - 248\*x) \* \Hh(-3,0)
  \nn\\[-0.1mm]&& \mbox{}
       + (368/3 - 424/3\*x^{-1} - 352/3\*x + 32/3\*x^2) \* \Hhh(1,1,0)
       + (160 + 352/3\*x^{-1} + 80\*x + 64/3\*x^2) \* \Hh(-1,2)
  \nn\\[-0.1mm]&& \mbox{}
       + (240 - 96\*x^{-1} - 136\*x) \* \Hhh(-2,0,0)
       + (1160/9 - 4136/9\*x^{-1} - 604/9\*x + 760/3\*x^2) \* \Hh(1,0)
  \nn\\[-0.1mm]&& \mbox{}
       + (64 - 256\*x^{-1} - 176\*x) \* \Hh(2,2)
       + (352 + 416\*x^{-1} + 240\*x) \* \H(0)\*\z3
       + (232 - 160\*x^{-1} - 76\*x) \* \Hhh(2,0,0)
  \nn\\[-0.1mm]&& \mbox{}
       + (192 - 248\*x^{-1} - 72\*x + 32\*x^2) \* \Hh(1,2)
       + (6896/9 + 4664/9\*x^{-1} + 4220/9\*x + 1552/9\*x^2) \* \H(2)
  \nn\\[-0.1mm]&& \mbox{}
       + (808/9 - 532/3\*x^{-1} - 440/9\*x + 560/9\*x^2) \* \Hh(1,1)
       - 8\*\pgq( - x) \* (
       4\*\H(-2,2) 
       - \H(-1)\*\z3 
       + 22\*\Hhh(-1,-2,0) 
  \nn\\[-0.1mm]&& \mbox{}
       - 2\*\Hh(-1,-1)\*\z2 
       - 12\*\Hhhh(-1,-1,-1,0) 
       + 12\*\Hhhh(-1,-1,0,0) 
       - 4\*\Hhh(-1,-1,2) 
       + 10\*\Hh(-1,0)\*\z2 
       - 21\*\Hhhh(-1,0,0,0) 
  \nn\\[-0.1mm]&& \mbox{}
       - 12\*\Hhh(-1,2,0) 
       - 4\*\Hhh(-1,2,1) 
       + 4\*\Hh(-1,3) 
       )
       + 8\*\pgq(x) \* ( 
       41\*\H(1)\*\z3 
       - 6\*\Hhh(1,-2,0) 
       - 8\*\Hh(1,0)\*\z2 
       + 35\*\Hhhh(1,0,0,0) 
  \nn\\[-0.1mm]&& \mbox{} 
       + 6\*\Hh(1,1)\*\z2 
       + 10\*\Hhhh(1,1,0,0)
       + 12\*\Hhhh(1,1,1,0) 
       + 4\*\Hhhh(1,1,1,1) 
       + 8\*\Hhh(1,1,2) 
       + 40\*\Hhh(1,2,0) 
       + 4\*\Hhh(1,2,1) 
       + 10\*\Hh(1,3) 
       )
          \big)
  \nn\\[1mm]&& \mbox{\hspn}
+\colour4colour{\cfs\*\nf}  \*  \big(
       - 12803/27 + 14008/81\*x^{-1} + 30475/54\*x - 21784/81\*x^2 
       + (64 - 32\*x) \* \Hhhh(0,0,0,0)
  \nn\\[-0.1mm]&& \mbox{}
       - ( 204 + 128/3\*x^{-1} - 142/3\*x + 256/9\*x^2) \* \Hh(0,0)
       - ( 176/3 + 128/3\*x^{-1} - 160/3\*x) \* \Hhh(0,0,0)
  \nn\\[-0.1mm]&& \mbox{}
       - ( 484/27 - 448/27\*x^{-1} - 464/27\*x) \* \H(1)
       + (160/9 - 160/9\*x^{-1} - 104/9\*x) \* (\Hh(1,1) - 2\*\H(2))
  \nn\\[-0.1mm]&& \mbox{}
       - ( 2330/27 - 200/9\*x^{-1} - 2494/27\*x - 3040/27\*x^2) \* \H(0)
       - 8/3\*\pgq(x) \* (
       3\*\H(1)\*\z2 
       + 5\*\Hh(1,0) 
  \nn\\[-0.1mm]&& \mbox{}
       + 5\*\Hhh(1,0,0)
       + \Hhh(1,1,0) 
       - \Hhh(1,1,1) 
       + \Hh(1,2) 
       + 2\*\Hh(2,1) 
       - 4\*\H(3) 
       )
          \big)
  \nn\\[1mm]&& \mbox{\hspn}
+\colour4colour{\ca\*\cf\*\nf}  \*  \big(
         332/27 + 3778/81\*x^{-1} - 2624/27\*x + 1892/81\*x^2
       + (64 - 80\*x^{-1} - 40\*x) \* \z3
  \nn\\[-0.1mm]&& \mbox{}
       - ( 256/9 - 656/9\*x^{-1} + 464/9\*x + 128/9\*x^2) \* \Hh(0,0)
       - ( 160/9 + 160/9\*x^{-1} + 104/9\*x) \* \Hh(-1,0)
  \nn\\[-0.1mm]&& \mbox{}
       - ( 160/9 - 160/9\*x^{-1} - 104/9\*x) \* \Hh(1,1)
       + (32/3 + 128/3\*x^{-1} + 128/3\*x) \* \Hhh(0,0,0)
  \nn\\[-0.1mm]&& \mbox{}
       + (136/9 - 160/9\*x^{-1} - 104/9\*x) \* \H(2)
       + (448/27 - 448/27\*x^{-1} - 392/27\*x) \* \H(1)
  \nn\\[-0.1mm]&& \mbox{}
       + (1280/27 + 1156/27\*x^{-1} + 308/27\*x + 64/9\*x^2) \* \H(0)
       - ( 320/9 - 320/9\*x^{-1} - 136/9\*x) \* \Hh(1,0)
  \nn\\[-0.1mm]&& \mbox{}
       - (32/3\*x^{-1} + 16/3\*x) \* \H(0)\*\z2
       + (8/3 - 160/9\*x^{-1} - 176/9\*x) \* \z2
       + 16/3\*\pgq( - x) \* (
       \Hh(-2,0) 
  \nn\\[-0.1mm]&& \mbox{}
       + \Hhh(-1,0,0) 
       )
       + 8/3\*\pgq(x) \* (
        6\*\H(1)\*\z2 
       + 10\*\Hhh(1,0,0) 
       + 2\*\Hhh(1,1,0) 
       - \Hhh(1,1,1) 
       + \Hh(2,1) 
       - 2\*\H(3) 
       )
          \big)
%%;
%%STOP
\:\: .
\end{eqnarray}

%\vspace{-4mm}
Unlike their spacelike counterparts, the off-diagonal timelike splitting
functions show a double-logarithmic enhancement of higher-order terms not only
for $x \ra 1$, but also for $x \ra 0$ \cite{LLxto0} with
\bea
\label{eq:PTqg1to1} 
 P^{\,(1)T}_{\rm qg}(x) &\!\! =\!\! &
%%START
%%L %%texPqg1tx1 =
   4\,\* \caf\* \nf\, \* L_1^2  
   \: + \: \left[ \,
       \frak{44}{3}\: \* \caf
     + \frak{8}{3}\: \* (\cf - \nf)
     \right] \,\* \nf\, \* L_1
%%;
%%STOP
  \; + \: {\cal O}(1)
  \: , \:
\\[1mm]
\label{eq:PTgq1to1}
 P^{\,(1)T}_{\rm gq}(x) &\!\! =\!\! &
%%START
%%L %%texPgq1tx1 =
   -\,4\, \* \caf\* \cf\, \* L_1^2
   \: - \: 8\, \* \caf\* \cf\, \* L_1
%;
%%STOP
  \; + \: {\cal O}(1)
\eea
 
\vspace*{-4mm}
\nin
and

\vspace*{-10mm}
\bea
\label{eq:PTqg1to0}
 xP^{\,(1)T}_{\rm qg}(x) &\!\! =\!\! &
%%START
%%L %%texPqg1tx0 =
       -\, \frak{80}{9}\: \* \ca\* \nf
%%;
%%STOP
  \; + \: {\cal O}(xL_0^2)
  \: , \:
\\[1mm]
\label{eq:PTgq1to0}
 xP^{\,(1)T}_{\rm gq}(x) &\!\! =\!\! &
%%START
%%L %%texPgq1tx0 =
   - 16\,\* \cf\*\ca\, \* L_0^2 
   \:-\: 24\,\* \cf\*\ca\, \* L_0 \:+\: \frak{68}{9}\: \* \cf\*\ca
%;
%%STOP
  \; + \: {\cal O}(xL_0^2) \:\: . \;\; 
\eea
Here and below we use the abbreviations
$\,L_1 = -\,\H(1)(x) = \ln \x1$, $\,L_0 = \H(0)(x) = \ln x\,$ and 
$\,\caf = C_A - C_F$.
The large-$x$ and small-$x$ limits of the new result (\ref{eq:PTqg2}) and 
(\ref{eq:PTgq2}) are given by 
\bea
\label{eq:PTqg2to1} 
  P^{\,(2)T}_{\rm qg}(x) &\!\! =\!\! &
%%START
%%L %%texPqg2tx1 =
   \frak{4}{3}\: \* \cafs\* \nf\, \* L_1^4
   \: + \: \left[ \,
       \frak{110}{9}\: \* \cafs
     + \frak{20}{9}\: \* \caf \* (\cf - \nf)
     \right] \,\* \nf\, \* L_1^3
  \nn \\[1.5mm]&& \mbox{\hspn\hspn}
   + \,\left[ \,
       \left( \, \frak{631}{9} - 8\,\*\z2 \right) \* \cafs
     + \left( \, \frak{652}{9} - 16\,\*\z2 \right) \* \caf \* \cf
     - \frak{172}{9}\: \*\caf \* \nf 
     + \frak{4}{3}\: \* (\cf - \nf)^2
     \right] \,\* \nf\, \*  L_1^2
  \nn\\[1.5mm]&& \mbox{\hspn\hspn}
   + \,\left[ \,
       \left( \, \frak{4156}{27} - \frak{176}{3}\,\*\z2 
             + 16\,\*\z3 \right) \* \cafs
     + \left( \, \frak{5914}{27} - 84\,\*\z2 + 96\,\*\z3 \right) \* \caf\*\cf
     + \frak{424}{9}\: \*\cfs 
  %%
  %%
%%STOP
 \right.  \\[1.5mm]&& \mbox{\hspn} \left.
%%START
     - \left( \, \frak{1672}{27} - \frak{32}{3}\,\*\z2 \right) \* \caf\*\nf
     - \frak{392}{9}\: \*\cf\*\nf 
     + \frak{40}{9}\: \*\nfs 
     - \frak{40}{3}\:\*\z2\,\* \cf\* (\cf - \nf) 
     \right] \,\* \nf\, \* L_1
%%;
%%STOP
  \; + \: {\cal O}(1)
  \: , \:
\nn \\[2mm]
\label{eq:PTgq2to1}
  P^{\,(2)T}_{\rm gq}(x) &\!\! =\!\! &
%%START
%%L %%texPgq2tx1 =
   \frak{4}{3}\: \* \cf\* \cafs\, \* L_1^4
   \: + \: \left[ \,
       \frak{50}{9}\: \* \cafs
     - \frak{4}{9}\: \* \caf \* (\cf - \nf)
     \right] \,\* \cf\, \* L_1^3
  \nn\\[1.5mm]&& \mbox{\hspn\hspn}
   + \left[ \,
       \frak{52}{9}\: \*\caf \* \nf
     - \left( \, \frak{334}{9} - 8\,\*\z2 \right) \* \cafs
     - \left( \, \frak{640}{9} - 16\,\*\z2 \right) \* \caf \* \cf
     \right] \,\* \cf\, \*  L_1^2
  \nn\\[1.5mm]&& \mbox{\hspn\hspn}
   + \left[ \, 
       \left( -\, \frak{2774}{27} + \frak{8}{3}\,\*\z2
             + 80\,\*\z3 \right) \* \cafs
     - \left( \, \frak{2360}{27} - \frak{148}{3}\,\*\z2 + 48\,\*\z3 \right) 
             \* \caf\*\cf
  %%
  %%
%%STOP
 \right.  \\[1.5mm]&& \mbox{\hspn} \left.
%%START
     + \left( \, \frak{392}{27} - \frak{32}{3}\,\*\z2 \right) \* \caf\*\nf
     + \frak{4}{3}\: \* (1+2\,\*\z2)\,\* \cf\,\* (\cf - \nf)
     \right] \,\* \cf\, \* L_1
%%;
%%STOP
  \; + \: {\cal O}(1)
  \:\: ,
\eea
%
%and
%
\bea
\label{eq:PTqg2to0}
 xP^{\,(2)T}_{\rm qg}(x) &\! =\! &
%%START
%%L %%texPqg2tx0 =
   - \frak{64}{9}\: \* \cas\* \nf\, \* L_0^3
   \: - \: \left[ \,
       \frak{64}{9}\: \* \cas\* \nf
     + \frak{32}{9}\: \* (\ca - 2\,\*\cf)\, \* \nfs
     \right] \,\* L_0^2
  \nn\\[1.5mm]&& \mbox{\hspn\hspn}
   + \left[ 
       \left( 40 + \frak{64}{3}\,\*\z2 \right) \* \cas\* \nf 
     - \frak{256}{27}\, \* (\ca - 2\,\*\cf)\, \* \nfs
     \right] \,\* L_0
%%;
%%STOP
  \; + \: {\cal O}(1)
  \:\: ,
\\[2mm]
\label{eq:PTgq2to0}
 xP^{\,(2)T}_{\rm gq}(x) &\! =\! &
%%START
%%L %%texPgq2tx0 =
   \frak{64}{3}\: \* \cf\* \cas\, \* L_0^4
   \: + \: \left[ \,
        \frak{928}{9}\: \* \cf\* \cas
      + \frak{64}{9}\: \* \cf \* \nf \* (\ca - \cf)
      \right] \,\* L_0^3
  \nn\\[1.5mm]&& \mbox{\hspn\hspn}
   + \left[
       \big( 40 - 64\,\* \z2 \big)\, \* \cfs\* \ca 
     + \left( \frak{1960}{9} - 64\,\* \z2 \right) \* \cf\* \cas
     - \frak{64}{3}\: \* \cfs \* \nf
     + \frak{328}{9}\: \* \cf \* \ca \* \nf
     \right] \,\* L_0^2
  \nn\\[1.5mm]&& \mbox{\hspn\hspn}
   + \left[
       \left( \frak{13384}{27} - \frak{752}{3}\,\* \z2 
             + 416\,\*\z3 \right) \* \cf\* \cas
     \:+\: \big( 244 + 192\,\* \z2 - 832\,\*\z3 \big)\, \* \cfs\* \ca
  %%
  %%
%%STOP
 \right.  \\[1.5mm]&& \mbox{\hspn} \left.
%%START
     + \big( 208 - 192\,\* \z2 + 128\,\*\z3 \big)\, \* \cft
     + \left( \frak{1156}{27} - \frak{32}{3}\,\* \z2 \right) \* \cf\* \ca\* \nf
     + \frak{200}{9}\: \* \cfs\* \nf  
     \right] \,\* L_0
%%;
%%STOP  
  \; + \: {\cal O}(1)
  \:\: . \nn
\eea

The leading logarithms in all four large-$x$ limits above are identical to 
those of the corresponding spacelike splitting functions, in agreement with the 
all-order prediction in Ref.~\cite{AV2010}.
As in the spacelike case \cite{MVV4},
all double-logarithmic contributions to these equations vanish for $\ca = \cf$.
Only one coefficient, that of $\ln^1 \x1$ in Eq.~(\ref{eq:PTqg2to1}), does not 
vanish in the supersymmetric limit $C_A = C_F = n_f$. 
Also this feature is 
analogous to the spacelike case discussed in Ref.~\cite{MVV4}. 
Eq.~(\ref{eq:PTgq2to1}) is in 
complete agreement with the corresponding result of Ref.~\cite{Grunb11}, see
also Ref.~\cite{BK06}.
This would not be the case if a term with $\,\ln \x1$ were present on the
right-hand-side of Eq.~(\ref{Pgq2AC}). 
 
\vspace{-0.5mm}
The coefficients of the leading small-$x$ logarithms of $P_{\rm qg}^{\,T}$ and
$P_{\rm gq}^{\,T}$ at NLO and NNLO are larger and smaller, respectively, by a 
factor $C_A/C_F$ than those of $P_{\rm qq}^{\,T}$ and $P_{\rm gg}^{\,T}$ in 
Eqs.~(13) -- (15) of Ref.~\cite{MV2}. For $P_{\rm gq}^{\,T}$ and 
$P_{\rm gg}^{\,T}$ this relation and the corresponding coefficients have been 
derived to all orders in Refs.~\cite{LLxto0}. Unlike at NLO, $P_{\rm qq}^{\,T}$ 
and $P_{\rm qg}^{\,T}$ are suppressed by only one power of $\ln \x1$ relative 
to $P_{\rm gq}^{\,T}$ and $P_{\rm gg}^{\,T}$ at NNLO, and presumably some or 
all higher orders. 

\vspace{-0.5mm}
We now turn to the additional constraints mentioned above Eq.~(\ref{Pqg2AC}).
These are provided by the well-known supersymmetric relations 
for the choice $C_A \,=\, C_F \, =\, \nf$ of the colour factors leading to a 
${\cal N}\! =\! 1$ supersymmetric theory. 
On the one hand, we can investigate the combinations
\beq
\label{susyST}
  \Delta_{\,\rm A}^{(n)}(x) \;=\;
  P_{\rm qq}^{(n)A}(x) + P_{\rm gq}^{(n)A}(x) -
  P_{\rm qg}^{(n)A}(x) - P_{\rm gg}^{(n)A}(x) \quad \mbox{ with } \quad
  A \,=\, S,\: T
\eeq
which vanish at LO, while at NLO and $x < 1$ these quantities are given by 
\bea
\label{susy1s}
  \Delta_{\,\rm S}^{(1)}(x) &\! =\!\! &
  \phantom{-\,} \frak{8}{3}\:\* x^{\,-1}
  \,+\, ( 4 - 8\,\*x - 16\,\* x^2)\,\* \ln x 
  \,+\, \frak{10}{3} - \frak{92}{3}\:\* x + 28\,\* x^2 
%  \,-\, 2\,\*\delta \x1 
  \; ,
\nn \\[1mm]
\label{susy1t}
  \Delta_{\,\rm T}^{(1)}(x) &\! =\!\! &
  -\, \frak{8}{3}\:\* x^{\,-1}
  \,-\, ( 4 - 8\,\*x - 16\,\* x^2)\,\* \ln x 
  \,+\, \frak{26}{3} + \frak{20}{3}\:\* x - 4\,\*x^2
%  \,-\, 2\,\*\delta \x1 
\eea
in the \MSb\ scheme \cite{FP80}, see also Ref.~\cite{AF81}.
Obviously $\Delta_{\,\rm S}^{(1)}\!$
and $\Delta_{\,\rm T}^{(1)}$ are much simpler than the individual NLO 
splitting functions. A further simplification is obtained by adding these two
quantities,
\beq
\label{dsusy1}
  \Delta_{\,\rm S}^{(1)}(x) + \Delta_{\,\rm T}^{(1)}(x) \;=\;
  12 - 24\,\*x + 24\,\*x^{\,2} \;=\; 12\,\* \pqg(x) \; ,
\eeq
where $\pqg(x)$ has been defined in Eq.~(\ref{eq:PTqg1}).
Using the results of Refs.~\cite{MVV3,MVV4,MMV06,MV2} and Eqs.~(\ref{eq:PTqg2})
and (\ref{eq:PTgq2}) the corresponding NNLO quantities are found to be 
(also at $x < 1$)
\bea
\label{susy2s}
  \Delta_{\,\rm S}^{(2)}(x) &\!\! =\!\! &
  -\,2\,\* \ln^3 x \,-\, 9\,\*\ln^2 x 
  \,-\, \left(\, \frak{8}{3}\: x^{\,-1} + \frak{368}{9} + 24\,\*\z2 \right)
        \,\* \ln x
  \;+\; \ldots
  \;+\; 8\,\* \ln \x1
\\[1mm]
  \Delta_{\,\rm T}^{(2)}(x) &\!\! =\!\! &
  -\,2\,\* \ln^3 x \,+\, \big(\, 16\,\* x^{\,-1} - 21 \big)\,\*\ln^2 x
  \,+\, \left(\, \frak{928}{9}\: x^{\,-1} + \frak{388}{9} + 24\,\*\z2 \right)
        \,\* \ln x
  \;+\; \ldots
  \;+\; 8\,\* \ln \x1 \:\: ,
\nn 
\label{susy2t}
\eea
where we have suppressed all contributions which are regular for $x \!\ra\! 0$
 and $x \!\ra\! 1$ for brevity,~and
\beq
\label{dsusy2}
  \Delta_{\,\rm S}^{(2)}(x) + \Delta_{\,\rm T}^{(2)}(x) \;=\;
  -\,24\,\*\z2\,\* \pqg(x) \:+\: \mbox{non-$\z2$ terms} \:\: .
\eeq
The latter means the absence of $\z2$ in the 
expansion about $x=1$ to all orders in $\x1$, cf.~Ref.~\cite{MV5}.
 
\vspace{-0.5mm}
The absence of $\z2\, x^{\,-1} \ln x$ also in the second line of 
Eq.~(\ref{susy2s}) provides a check of the coefficient of $\ln x$, and hence
(due to the second-moment constraint (\ref{Pgq2AC2})) of the whole $C_A$
coefficient in Eq.~(\ref{Pgq2AC}). An additional $\ln x$ term in 
Eq.~(\ref{Pqg2AC}), except with a prefactor $C_A\! -\!C_F$, would spoil the
symmetry between the two lines of Eq.~(\ref{susy2s}) and hence also Eq.~%
(\ref{dsusy2}). Obviously non-$(C_F\!-\!C_A)$ terms not proportional to 
$\pgq(x)$ on the r.h.s.~of Eq.~(\ref{Pqg2AC}) would also conflict with the 
latter relation.

A second aspect concerns the analytic structure of the physical evolution kernels, 
in particular the differences between the analytic continuations ($x \ra 1/x$) of the spacelike and timelike ones.
From Eq.~(\ref{eq:ACphys1}) we define the matrix $\Delta K^{(n)}$ as
\beq
\label{eq:dACK}
  \Delta K^{(n)}(x)  \,=\, \mbox{\it AC}\: \big[ K^{(n)S}(x) \big] - K^{(n)T}(x)
  \, ,
\eeq
with the restriction $x < 1$ and 
entries according to Eqs.~(\ref{eq:ACphys2}), (\ref{Pdiag2AC}), (\ref{Pgq2AC}) and (\ref{Pqg2AC}).
It is interesting to investigate whether Eq.~(\ref{eq:dACK}) directly respects Gribov-Lipatov reciprocity.
In Mellin space, this feature implies that the corresponding expressions are functions only of
the product $N(N+1)$ of the Mellin variable $N$, i.e., parity preserving, 
a fact already exploited in the large-$x$ (large-$N$) limit of Ref.~\cite{Grunb11}.
The eigenvalues $\lambda_{i}$, $i=1,2$ of $\Delta K^{(n)}$ are determined from the characteristic equation
\beq
\label{eq:eigenvals-dACK}
  \lambda_{i}^2 - \lambda_{i}\, {\rm tr}\left(\Delta K\right) + {\rm det}\left(\Delta K\right) 
  \;=\; 0
  \, ,
\eeq
and, following~\cite{BK06}, it is sufficient to study the conditions which 
Gribov-Lipatov reciprocity imposes on the trace and determinant of $\Delta K^{(n)}$ in Eq.~(\ref{eq:dACK}), 
i.e.,
\bea
\label{eq:tr-dACK}
AC \left[{\rm tr}\left(\Delta K^{(n)}(x)\right)\right] 
 ~~ - ~~ \:{\rm tr}\left(\Delta K^{(n)}(x)\right)
  &\!=\!& 0 
  \; ,
\\[1mm]
\label{eq:det-dACK}
AC \left[{\rm det}\left(\Delta K^{(n)}(x)\right)\right] 
- \:{\rm det}\left(\Delta K^{(n)}(x)\right)
  &\!=\!& 0 
  \; .
\eea
%In QCD, Eq.~(\ref{eq:tr-dACK}) is fulfilled to NNLO due to Eq.~(\ref{eq:ACphys2}), 
In QCD, Eq.~(\ref{eq:tr-dACK}) is fulfilled to NNLO due to Eq.~(\ref{Pdiag2AC}), 
while Eq.~(\ref{eq:det-dACK}) is not. 
However, in the supersymmetric limit, $C_A \,=\, C_F \, =\, \nf$, 
also Eq.~(\ref{eq:det-dACK}) holds to NNLO 
%as a result of non-trivial relations between the colour factors in Eqs.~(\ref{Pgq2AC}) and (\ref{Pqg2AC}).
as a result of non-trivial relations between the coefficients in Eqs.~(\ref{Pgq2AC}) and (\ref{Pqg2AC}).
Eq.~(\ref{eq:det-dACK}) provides thus a further constraint on $P_{\rm qg}^{\,(2)T}$ and $P_{\rm gq}^{\,(2)T}$,  
again except for contributions proportional to $C_A\! -\!C_F$, which vanish 
trivially in the transition to a ${\cal N}\! =\! 1$ supersymmetric theory.

\begin{figure}[p]
\vspace*{-1mm}
\centerline{\epsfig{file=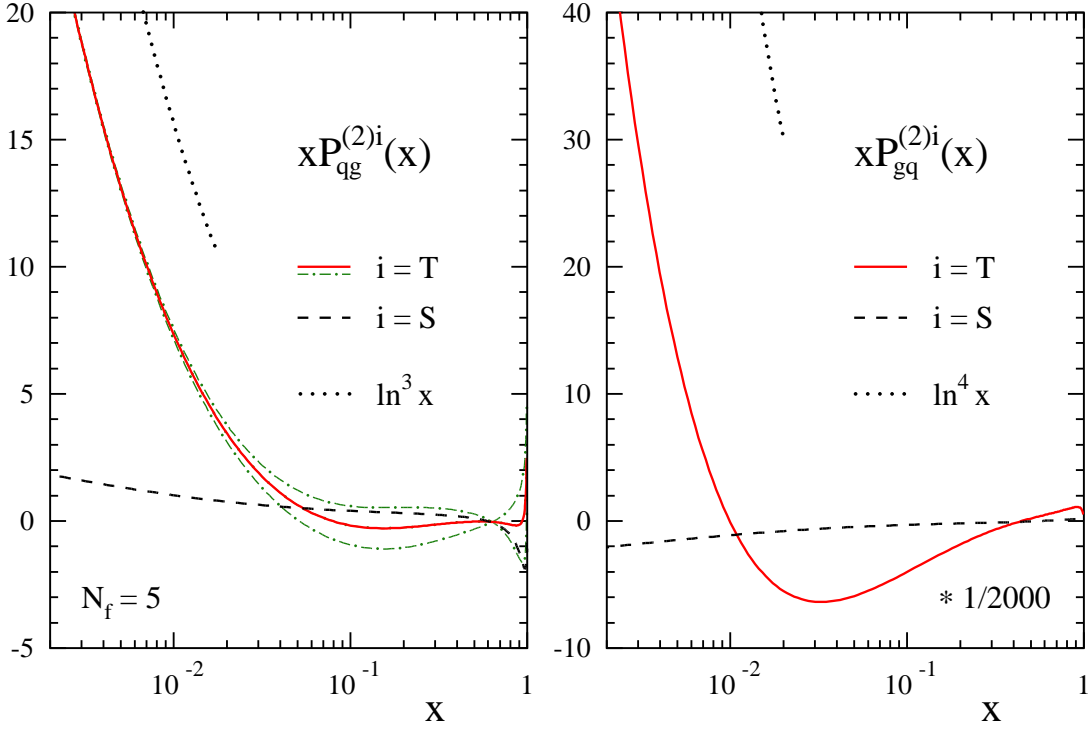,width=14.8cm,angle=0}}
\vspace{-3mm}
\caption{ \label{pic:P2tsx0}
 The third-order timelike off-diagonal (quark-gluon and gluon-quark) splitting
 functions for five flavours, multiplied by $x$ and divided by $2000 \simeq
 (4\pi)^3$ for display purposes.
 The remaining uncertainty of the former quantity is indicated by the
 dash-dotted curves.
 Also shown are the respective leading small-$x$ contributions and the
 corresponding spacelike splitting functions.}
\vspace{1mm}
\end{figure}
\begin{figure}[p]
\vspace{-2mm}
\centerline{\epsfig{file=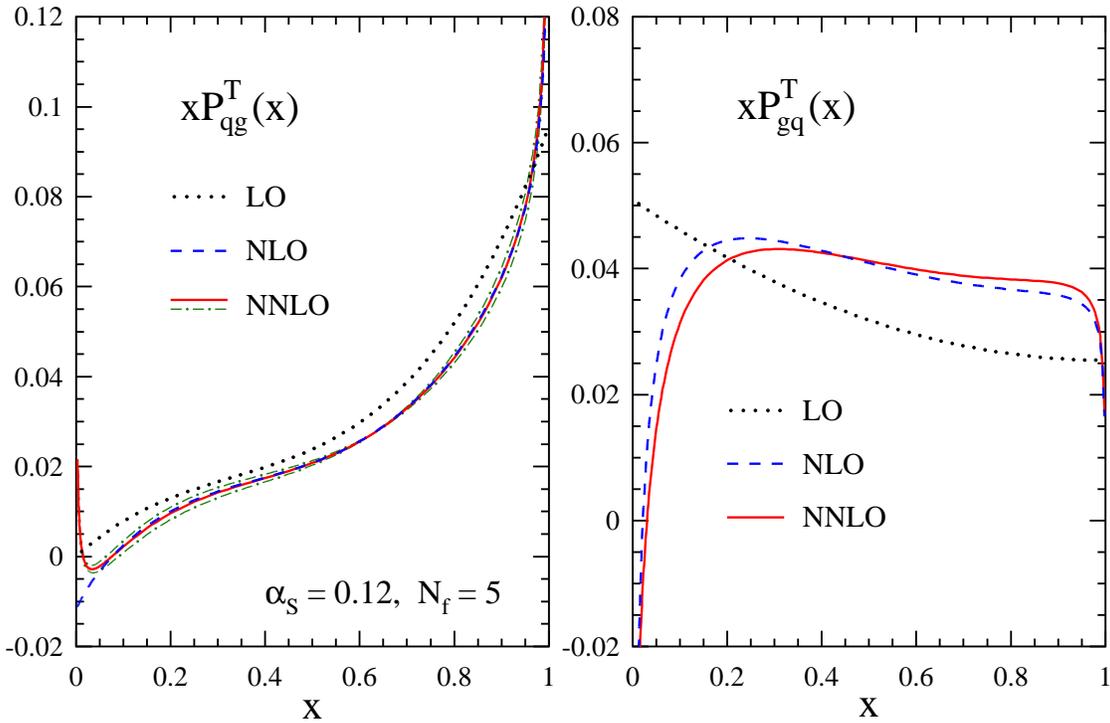,width=14.8cm,angle=0}}
\vspace{-3mm}
\caption{ \label{pic:Ptexp}
 The resulting perturbative expansion of the timelike quark-gluon and
 gluon-quark splitting functions, again multiplied by $x$, at a typical value
 of the strong coupling constant.}
\vspace{-2mm}
\end{figure}

In summary, these considerations are still not sufficient to definitely fix the
right-hand-side of Eq.~(\ref{Pqg2AC}). As an estimate of the remaining
uncertainty we suggest to use the offset
\beq
\label{eq:dPTqg2}
 \delta P_{\rm qg}^{\,(2)T}(x) \:\: = \:\: \pm \,
 2\,\*\z2\,\*\b0\* (C_A-C_F)\,\* ( 11 + 24\,\*\ln x )\,\* P_{\rm qg}^{\,(0)T}(x)
 \; .
\eeq
The functions (\ref{eq:PTqg2}) and (\ref{eq:PTgq2}), the former including the
error band due to Eq.~(\ref{eq:dPTqg2}), are shown and compared to their
spacelike counterparts in Fig.~1; and the LO, NLO and NNLO approximations to
Eq.~(\ref{eq:PTexp}) are illustrated in Fig.~2 at the typical scale $\Qs \simeq
M_Z^{\,2}$.
As in the diagonal cases \cite{MV2}, the higher-order corrections are much
larger at small values of $x$ than in the spacelike case.  The \mbox{small-$x$}
behaviour of $P_{\rm qg}^{\,(2)T}$ and $P_{\rm gq}^{\,(2)T}$ is
similar to that of $P_{\rm qq}^{\,(2)T}$ and $P_{\rm gg}^{\,(2)T}\!$,
respectively, with particularly large cancellations between the powers of
$\ln x$ occurring in $P_{\rm gq}^{\,(2)T}$ and $P_{\rm gg}^{\,(2)T}\!$.

For the use of the NNLO splitting functions in numerical analyses we have 
prepared, analogous to Refs.~\cite{MVV3,MVV4}, compact and accurate 
parametrizations not only of the present results, but also of the non-singlet
and diagonal quantities derived in Refs.~\cite{MMV06,MV2}. 
These parametrizations can be found in Appendix B. 
Corresponding {\sc Fortran} files, and {\sc Form} files of our main results, 
can be obtained by downloading the source of this paper from the {\tt arXiv} 
servers or from the authors upon request. 
This includes the rather lengthy (even or odd) integer-$N$ Mellin-space expressions in terms of
harmonic sums \cite{Hsums} which we have not presented here for brevity.

To summarize, we have performed an indirect determination of the hitherto 
unknown off-diagonal NNLO timelike splitting functions (\ref{eq:PTqg2}) and
(\ref{eq:PTgq2}) for the evolution of parton-to-hadron fragmentation functions 
$\,D_{\! f}^{\,h}(x,\Qs)$. 
We expect the remaining uncertainty of the former quantity, estimated in 
Eq.~(\ref{eq:dPTqg2}) and illustrated in Figs.~1 and 2, to be phenomenologically 
acceptable. 
Hence, combining these results with those of Refs.~\cite{MMV06,MV2,RvN96}, NNLO analyses are now possible of data on the transverse fragmentation 
function (but not yet its longitudinal counterpart, where also the third-order 
coefficient functions are required) in semi-inclusive electron-positron 
annihilation.

The remaining uncertainty of $P_{\rm qg}^{\,(2)T}\!$ does not affect the
logarithmically enhanced large-$x$ and small-$x$ contributions. We expect that
these results can be useful to improve the corresponding resummations. In fact,
an extension of the generalized large-$x$ resummation of Ref.~\cite{ASV1} to the
present timelike case will be presented in a forthcoming publication.

Further research is required to completely fix $P_{\rm qg}^{\,(2)T}\!$ and 
to check our result for $P_{\rm gq}^{\,(2)T}\!$. A direct calculation of the
leading-$\nf$ contribution to the former quantity should be possible, but would
not address the critical contributions, while a full $x$-dependent diagram 
calculation beyond these terms appears to remain formidable task. 
A computation of 
the $N=4$ and $N=6$ moments of the $\nfs$ and $\nf$ parts, respectively, of 
Eqs.~(\ref{eq:PTqg2}) and (\ref{eq:PTgq2}) presumably would be sufficient and 
may be feasible, e.g., generalizing the approach used in Ref.~\cite{MM06}, in 
the foreseeable future.

\vspace*{6mm}
\nin
{\bf Acknowledgements:}
The work of A.A. and A.V has been supported by the UK Science \& Technology
Facilities Council (STFC) under grant number ST/G00062X/1, and that of S.M.
by the Deutsche Forschungsgemeinschaft in Sonderforschungsbereich/Transregio~9.
We are also part of the European-Union funded network {\it LHCPhenoNet},
contract number PITN-GA-2010-264564.
Our analytic calculations were performed using {\sc Form} \cite{Form3}, and the 
numerical evaluation of the harmonic polylogarithms was carried out using the
program of Ref.~\cite{HPLnum}.

\renewcommand{\theequation}{A.\arabic{equation}}
\setcounter{equation}{0}

\vspace{2mm}
\subsection*{Appendix A}
The first- and second-order coefficient functions for SIA via an intermediate
scalar in Eq.~(\ref{FKCmat})~read
\bea
\label{app:cphig1t}
 c_{\phi,\rm g}^{\,(1)}(x) &\! =\! &
%%START
%%L %%texcphig1t =
 \colour4colour{\ca} \,\* \big(
       11/3\,\* ( 1 + x + x^2 - x_1^{\,-1} )
     + \pgg(x) \* ( 8\,\*\H(0) - 4\,\*\H(1) ) 
     + ( 67/9 + 8\,\*\z2 )\, \* \delta(1-x)
 \big)
  \nn\\ & & \mbox{\hspn}
 - \, \colour4colour{\nf} \,\*  \big(
       2/3\,\* ( 1 + x + x^2 - x_1^{\,-1} )
     + 10/9\,\* \delta(1-x)
 \big)      
%%;
%%STOP
\; ,
\\
\label{app:cphiq1t}
 c_{\phi,\rm q}^{\,(1)}(x) &\! =\! &
%%START
%%L %%texcphiq1t =
 \colour4colour{\nf} \,\* \big(
     4\,\* x - 7\,\* x^2
     + \pqg(x) \* ( 4\,\*\H(0) - 2\,\*\H(1) )
 \big)
%%;
%%STOP
\eea
with $\pgg(x) = x^{\,-1} - 2 + x - x^2 + x_1^{\,-1}$, where
$f(x)\, x_1^{\,-1} \equiv f(x)/\x1$ has to be read as a $+$-distri\-bution if 
$f(x)$ does not vanish at $x=1$, and
\bea
\label{app:cphig2t}
 \lefteqn{c_{\phi,\rm g}^{\,(2)}(x) \:\: =\:\: 
%%START
%%L %%texcphig2t =
 \colour4colour{\cas}  \,\*  \big(
       (5099\,\* x^{\,-1} - 3301\,\*x^2 - 2570\,\*x_1^{\,-1})/27
     - 46/3 + 1858/9\,\*x 
     - ( 140/3 + 32/3\,\*x 
%%STOP
}
%%START
  \nn\\[-0.5mm]&& \mbox{}
       + 44/3\,\* (x^{\,-1} + 3\,\* x^2 - x_1^{\,-1} ))\,\*\z2
     - ( 101 - 2092\,\*x^{\,-1} - 2623\,\*x 
       - 268\,\* (2\,\* x^2 + 3\*x_1^{\,-1} ))/9\,\* \H(0)
  \nn\\&& \mbox{}
     - ( 162 + 138\,\*x + 484/3\,\*x^2 
       + 220/3\,\*( 2\,\*x^{\,-1} - x_1^{\,-1} )) \,\* \Hh(0,0)
     - ( 268 - 356\,\*x 
  \nn\\&& \mbox{}
       - 88\,\* ( 5\,\* x^{\,-1} - 2\,\* x^2 + x_1^{\,-1} ))/3 \,\* \H(2)
     + ( 160 - 46\,\*x + 340/3\,\*x^2 - 778/9\,\* ( x^{\,-1} + x_1^{\,-1} ))
       \,\* \H(1)
  \nn\\&& \mbox{}
     - 88/3\,\* ( 1 + x + x^2 - x_1^{\,-1} )\,\* \Hh(1,0)
     + ( 452 - 308\,\*x^{-1} - 232\,\*x 
       + 176\,\* ( 2\,\* x^2 - x_1^{-1} ))/3\,\* \Hh(1,1)
  \nn\\ & & \mbox{}
     + 8/3\,\* ( 6 + 11\,\* x^{\,-1} + 6\,\* x + 11\,\* x^2 )\,\* \Hh(-1,0)
     - 8\,\* (1+x)\,\* ( 
         8\,\*\z3 + 8\,\*\z2\,\* \H(0) + 22\,\* \Hhh(0,0,0) - 12\,\* \H(3) 
  \nn\\ & & \mbox{}
       + 4\,\* \H(2,1) 
       ) 
     + 4\*\pgg(-x)\,\* (
         7\,\*\z3 - 2\,\*\z2\,\* \H(0) + 9\,\* \Hhh(0,0,0) - 2\,\* \H(3)
       - 2\,\* \Hh(-2,0) - 8\,\*\z2\,\* \H(-1) + 4\,\* \Hh(-1,2)
  \nn\\ & & \mbox{}
       - 8\,\* \Hhh(-1,-1,0) - 6\,\* \Hhh(-1,0,0)
       )
     + 4\,\* \pgg(x)\,\* (
       - 31\,\*\z3 + 2\,\*\z2\,\* \H(0) + 6\,\* \Hh(-2,0) - 31\,\* \Hhh(0,0,0)
       + 2\,\*\H(3)
  \nn\\ & & \mbox{}
       - 2\,\*\z2\,\* \H(1) - 20\,\* \Hh(2,0) + 10\,\* \Hh(2,1) 
       + 6\,\* \Hhh(1,0,0) + 10\,\* \Hh(1,2) + 10\,\* \Hhh(1,1,0) 
       - 12\,\* \Hhh(1,1,1)
       )
  \nn\\ & & \mbox{}
     + \delta(1-x) \* (
         30425/162 - 242/3\,\*\z3 + 830/9\,\*\z2 + 101/5\,\*\zs2
        )
  \big)
  \nn\\[1mm]&& \mbox{\hspn}
 +\, \colour4colour{\ca\*\nf}  \,\*  \big(
       (358\,\* x^{\,-1} - 928 - 502\,\*x + 672\,\*x_1^{\,-1})/27 
     - 116/3\,\*x^2 
     + ( 362 - 260\,\*x^{\,-1} - 120\,\*x_1^{\,-1} 
  \nn\\[-0.5mm] && \mbox{}
       - 238\,\*x + 172\,\* x^2 )/9\,\* \H(0)
     + ( 36 - 16\,\* x^{\,-1} - 12\,\*x 
       + 40\,\*( x^2 - x_1^{\,-1} ))/3 \,\* \Hh(0,0)
     - ( 350 - 168\,\*x_1^{\,-1} 
  \nn\\&& \mbox{}
       - 88\,\*(2\,\* x^{\,-1} + x - 3\,\*x^2 ) )/9 \,\* \H(1)
     + 8/3\,\*( ( x^{\,-1} - 1 + 2\,\*x ) \* ( 2\,\* \Hh(1,0) - \Hh(1,1) )
     + ( 1 + x ) \* ( \H(2) - \z2 ) )
  \nn\\&& \mbox{}
     - 8/3\,\* \pgg(x)\,\* (\z2 + 2\,\*\Hh(1,0) - 4\,\*\Hh(1,1) + 2\,\*\H(2) )
     - \delta(1-x) \* ( 4112/81 + 8\,\*\z2 + 28/3\,\*\z3 )
     \big)
  \nn\\[1mm] & & \mbox{\hspn}
 +\, \colour4colour{\cf\*\nf}  \,\*  \big( 
       2\,\* x_1^{\,-1} 
       - (744 + 723\,\* x - 2017\,\* x^2 + 928\,\* x^{\,-1} )/27
     + (8 + 16/3\,\* (x^{\,-1} + x^2) + 12\,\* x )\,\* \z2 
  \nn\\[-0.5mm]&& \mbox{}
     - 8/9\,\* ( 63 + 19\,\* x^{\,-1} + 81\,\* x + 41\,\* x^2 )\,\* \H(0)
     + ( 14 + 22\,\* x + 32/3\,\*( x^{\,-1} + x^2))\,\* \Hh(0,0)
     - ( 12 + 8\,\* x
  \nn\\&& \mbox{}
       + 32/3\,\* x^{\,-1} )\,\* \H(2)
     + ( 78 - 16\,\* x^{\,-1} - 150\,\* x + 88\,\* x^2 )/9 \,\* \H(1)
     + 4/3\,\* ( 4\,\* x^{\,-1} - 3\,\* x - 4\,\* x^2 + 3 )\,\* \Hh(1,1)
  \nn\\&& \mbox{}
     + 4\,\* (1+x)\* (4\,\*\z3 + 4\,\*\z2\,\* \H(0) + 11\,\* \Hhh(0,0,0)
         - 6\,\* \H(3) + 2\,\* \Hh(2,1) )
     - \delta(1-x) \* ( 63/2 - 24\,\* \z3 )
     \big)
  \nn\\[1mm] & & \mbox{\hspn}
 +\, \colour4colour{\nfs}  \,\*  \big(
       8/27\,\* ( 1 + x + x^2 - x_1^{\,-1} ) \* ( 5 + 3\,\* \H(1) )     
     + \delta(1-x) \* ( 100/81 - 8/9\,\* \z2 )
     \big)
%%;
%%STOP
\:\: , \\[2mm]
\label{app:cphiq2t}
 \lefteqn{c_{\phi,\rm q}^{\,(2)}(x) \:\: =\:\: 
%%START
%%L %%texcphiq2t =
 \colour4colour{\ca\*\nf}  \,\*  \big(
     - (2590 + 296\,\* x^{\,-1} - 74\,\* x )/27 - 695/18\,\* x^2 
     + ( 22 - 60\,\* x + 56\,\* x^2 )\,\* \z2
%%STOP
}
%%START
  \nn\\[-0.5mm]& & \mbox{}
     + ( 16 + 88\, \* x )\, \* \z3
     + ( (196 + 48\,\* x^{\,-1} - 2810\,\* x + 582\,\* x^2)/9 + 16\,\* \z2 )
       \,\* \H(0)
     + ( 8 + 64\,\* x^{\,-1} 
  \nn\\&& \mbox{}
       + 212\,\* x + 116\,\* x^2 )/3\,\* \Hh(0,0)
     + ( 10 - 32\,\* x^{\,-1} - 212\,\* x + 180\,\* x^2 )/3 \,\* \H(2)
     + ( 44 + 248\,\* x )\,\* \Hhh(0,0,0)
  \nn\\&& \mbox{}
     - 8\,\* ( 1 + 4\,\* x + 2\,\* x^2 )\,\* \Hh(-2,0)
     - ( 44 + 72\,\* x + 40\,\* x^2 )\,\* \H(3)
     - 8/3\,\* ( 3 + 2\,\* x^{\,-1} + 9\,\* x + 8\,\* x^2 )\,\* \Hh(-1,0)
  \nn\\&& \mbox{}
     - ( 10/3 + 80\,\* x^2 - 4/3\,\* (4\,\* x^{\,-1} + 53\,\* x))\,\* \Hh(1,1)
     - ( 127 + 24\,\* x^{\,-1} - 818\,\* x + 1016\,\* x^2 )/9 \,\* \H(1)
  \nn\\&& \mbox{}
     - 8\,\* \Hh(1,0)
     + 8\,\* ( 4 - 2\,\* x + 6\,\* x^2 )\,\* \Hh(2,1)
     + 4\,\* \pqg(-x)\, \* ( \z2\,\* (3\,\* \H(0) - 2\,\*  \H(-1))
       - 5\,\* \Hhh(-1,0,0) + 2\,\* \Hh(-1,2) )
  \nn\\&& \mbox{}
     + \pqg(x)\, \* ( 8\*\z2\,\* \H(1) + 2/3\,\* \Hh(1,0) - 12\,\* \Hh(2,0)
       + 16\,\* \Hhh(1,0,0) + 8\,\* \Hh(1,2) + 8\,\* \Hhh(1,1,0)
       - 20\,\* \Hhh(1,1,1) )
     \big)
  \nn\\[1mm]&& \mbox{\hspn}
 + \,  \colour4colour{\cf\*\nf}  \,\*  \big(
     + 62 - 71\,\* x + 155/2\,\* x^2
     - ( 22 - 24\,\* x + 4\,\* x^2 )\,\* \z2
     - ( 40 - 112\,\* x + 96\,\* x^2 )\, \* \z3
  \nn\\[-0.5mm]& & \mbox{}
     - ( 10 - 47\,\* x + 98\,\* x^2 + 16\,\* \z2\,\* x^2 ) \,\* \H(0)
     - ( 7 + 80\,\* x - 104\,\* x^2 )\,\* \Hh(0,0)
     + 4\,\* ( 6 - 4\,\* x + x^2 )\,\* \H(2)
  \nn\\&& \mbox{}
     + 16\,\* ( 1 + x )\,\* \Hh(-1,0) 
     + 16\,\* ( 1 + 2\,\* x^2 )\,\* \Hh(-2,0) 
     + ( 1 + 36\,\* x - 48\,\* x^2 )\,\* \H(1)
     + ( 6 - 36\,\* x + 60\,\* x^2 )\,\* \Hh(1,0) 
  \nn\\&& \mbox{}
     - 4\,\* x\,\* \Hh(1,1)
     - x^2\,\* ( 
         44\,\* \Hhh(0,0,0) + 8\,\* \Hh(2,1) - 24\,\* \H(3) )
     - 8\,\* \pqg(-x)\,\* (
        \z2\,\* \H(-1) - \Hhh(-1,0,0) + 2\,\* \Hhh(-1,-1,0)  )
  \nn\\&& \mbox{}
     - \pqg(x)\,\* (
         4\,\* \z2\,\* ( \H(0) + 3\,\* \H(1) ) + 20\,\* \Hh(1,1)
       + 22\,\* \Hhh(0,0,0) + 28\,\* \Hh(2,0) - 8\,\* \H(3)
       + 4\,\* \Hhh(1,0,0) - 12\,\* \Hh(1,2) 
  \nn\\&& \mbox{}
       - 12\,\* \Hhh(1,1,0) + 4\,\* \Hhh(1,1,1) )
     \big) 
  \nn\\[1mm]&& \mbox{\hspn}
 + \,  \colour4colour{\nfs}  \,\*  \big(
       (112 - 392\,\* x + 833\,\* x^2)/27
     - 4/9\,\* ( 34 - 56\,\* x + 53\,\* x^2 )\,\* \H(0)
     + \pqg(x)\,\* (
         58/9\,\* \H(1) - 4\,\*\z2 
  \nn\\&& \mbox{}
      + 4/3\, \* ( \Hh(0,0) - \H(2) + \Hh(1,0) + \Hh(1,1) ) )
     \big)
%%;
%%STOP
\eea
As already discussed in Ref.~\cite{MV2}, the second moment of 
$c_{\phi,\rm g}^{\,(2)} + c_{\phi,\rm q}^{\,(2)}$ directly enters the NNLO
Higgs decay rate to hadrons in the heavy-top limit, and agrees with the result 
of Ref.~\cite{HDecay}, see also Ref.~\cite{HDecay2}.
We expect that these coefficient functions will be useful also for other 
theoretical studies. They can be employed, for instance, to extend the 
large-$x$ results of Ref.~\cite{Grunb11} to $P_{\rm qg}^{\,(2)T}$. 

\renewcommand{\theequation}{B.\arabic{equation}}
\setcounter{equation}{0}

\subsection*{Appendix B}
Since the exact expressions of the NNLO splitting functions are rather lengthy
and complex, it is useful to have at one's disposal compact but accurate
approximate representations which also can be transformed readily to Mellin 
space at all (complex) values of $N$. 
In this final appendix we therefore provide parametrizations of all NNLO 
timelike splitting functions in QCD which are built up, besides powers of $x$, 
only from the $+$-distribution and the end-point logarithms
$$
  \DD_{\,0} \: = \: 1/(1-x)_+ \: ,
  \quad L_1 \: = \: \ln (1-x) \: ,
  \quad L_0 \: = \: \ln x \:\: .
$$
  
The non-singlet splitting functions $P^{\,(2)T\pm}_{\,\rm ns}$ 
\cite{MMV06} can be represented~by 
\bea
\label{eq:P+appr}
  P^{(2)T+}_{\,\rm ns}(x)\!\!\! & \cong & \mbox{} \!\!
     + 1174.898\: \DD_0 + 1295.625\: \delta (1-x) - 707.67\: L_1
     + 1658.7 - 4249.4\: x 
  \nonumber \\[0.2mm] & & \mbox{} \!\!
     - 1075.3\: x^2 + 593.9\: x^3 - L_0 L_1 [56.907 + 519.37\: L_0 
     + 559.1\: x \, ] 
  \nonumber \\[0.2mm] & & \mbox{} \!\!     
     + 1327.5\: L_0 - 189.37\: L_0^2 - 352/9\: L_0^3 + 128/81\: L_0^4
  \nonumber \\[0.7mm] &+& \mbox{} \nf \:\big(
     - 183.187\: \DD_0 - 173.935\: \delta (1-x)
     + 5120/81\: L_1 - 198.1 + 466.29\: x 
  \nonumber \\[-0.5mm] & & \mbox{} \quad 
     + 181.18\: x^2 - 31.84\: x^3 - 39.113\: xL_0 
     - L_0 L_1 [50.758 - 85.72\: x 
  \nonumber \\[0.2mm] & & \mbox{} \quad 
     - 28.551\: L_0 + 23.102\: x L_0]
     - 168.89\: L_0 - 176/81\: L_0^2 + 64/27\: L_0^3\, \big) 
  \nonumber \\[0.7mm] &+& \mbox{} \n2f \:\big(
     - \DD_0 - (51/16 + 3\,\z3 - 5\,\z2) \: \delta (1-x) 
     + x\,(1-x)^{-1} L_0\, (3/2\: L_0 + 5) 
  \nonumber \\[-0.5mm] & & \mbox{} \quad
     + 1
     + (1-x)\, (6 + 11/2\: L_0 + 3/4\: L_0^2)\, \big) \:\, 64/81 
  \:\: 
\eea
and
\bea
\label{eq:P-appr}
  P^{(2)T-}_{\,\rm ns}(x)\!\!\! & \cong & \mbox{} \!\!
     + 1174.898\: \DD_0 + 1295.622\: \delta (1-x) - 707.94\: L_1
     + 1981.3 - 4885.7\: x 
  \nonumber \\[0.2mm] & & \mbox{} \!\!
     - 577.42\: x^2 + 407.89\: x^3 + L_0 L_1 [4563.2 + 1905.4\: L_0 
     - 5140.6\: x 
  \nonumber \\[0.2mm] & & \mbox{} \!\!
     + 1969.5\: x L_0] 
     - 34.683\: x L_0^4 - 437.03\: x L_0^3
     + 1625.5\: L_0 - 38.298\: L_0^2 
  \nonumber \\[0.2mm] & & \mbox{} \!\!
     - 1024/27\: L_0^3 - 140/81\: L_0^4
  \nonumber \\[0.7mm] &+& \mbox{} \nf \:\big(
     - 183.187\: \DD_0 - 173.9376\: \delta (1-x)
     + 5120/81\: L_1 - 217.84 + 511.92\: x 
  \nonumber \\[-0.5mm] & & \mbox{} \quad 
     + 209.19\: x^2 - 85.786\: x^3 + 92.453\: xL_0 
     + L_0 L_1 [71.428 + 30.554\: L_0 
  \nonumber \\[0.2mm] & & \mbox{} \quad 
     - 23.722\: x - 18.975\: x L_0]
     - 188.99\: L_0 - 784/81\: L_0^2 + 128/81\: L_0^3\, \big) 
  \nonumber \\[0.7mm] &+& \mbox{} \n2f \:\big(
     - \DD_0 - (51/16 + 3\,\z3 - 5\,\z2) \: \delta (1-x) 
     + x\,(1-x)^{-1} L_0\, (3/2\: L_0 + 5) 
  \nonumber \\[-0.5mm] & & \mbox{} \quad
     + 1
     + (1-x)\, (6 + 11/2\: L_0 + 3/4\: L_0^2)\, \big) \:\, 64/81 
  \:\: .
\eea
The $\n2f$ parts of $P^{(2)T\pm}_{\,\rm ns}$ (which are identical and equal 
to their spacelike counterparts in Ref.~\cite{MVV3}), the $+$-distribution 
contributions (up to a numerical truncation of the coefficients involving 
$\zeta_{i\,}$), and the rational coefficients of the (sub-)leading regular 
end-point terms are exact in Eqs.~(\ref{eq:P+appr}) and (\ref{eq:P-appr}). 
The remaining coefficients have been determined by fits to the exact results
at $10^{-6} \leq x \leq 1\! -\! 10^{-6}$, and finally the coefficients of
$\delta \x1$ have been adjusted very slightly using the lowest integer moments.
The difference between the NNLO `valence' and `minus' splitting functions is
equal to that in the spacelike case; a parametrization can be found in 
Eq.~(4.24) of Ref.~\cite{MVV3}).

\pagebreak
 
Corresponding representations of the pure-singlet and gluon-gluon
splitting functions \cite{MV2} are  
\bea
\label{eq:Pps-ap}
  P^{(2)T}_{\rm ps}(x)\!\!\! & \cong & \!\!\big\{ \: \nf \:
     \big( - 5.926\: L_1^3 - 9.751\: L_1^2 - 8.65\: L_1
     - 106.65 - 848.97\: x + 368.79\: x^2 
  \nn \\[-0.5mm] & & \mbox{} \quad     
     - 61.284\: x^3 + 96.171\: L_0 L_1
     + 656.49\: L_0 + 425.14\: L_0^2 + 47.322\: L_0^3 
  \nn \\[0.2mm] & & \mbox{} \quad   
     + 9.072\: L_0^4 + 479.87\: x^{\,-1}  
     + 324.07\: x^{\,-1} L_0  - 128/9\: x^{\,-1} L_0^2 
     - 256/9\: x^{\,-1} L_0^3 \, \big) 
  \nn \\[0.7mm] &+& \mbox{} \n2f \: \big(
     1.778\: L_1^2 + 16.611\: L_1 + 87.795 - 57.688\: x 
     - 41.827\: x^2 + 25.628\: x^3 
  \nn \\[-0.5mm] & & \mbox{} \quad
     - 7.9934\: x^4 - 2.1031\: L_0 L_1 
     + 26.294\: x L_0 - 7.8645\: x L_0^3
     + 57.713\: L_0 
  \nn \\[0.2mm] & & \mbox{} \quad
     + 9.1682\: L_0^2 - 1.9\: L_0^3 + 0.019122\: L_0^4 - 128/81\: x^{\,-1} 
     \, \big) \big\} (1-x)
\eea
and
\bea
\label{eq:Pgg-ap}
  P^{(2)T}_{\rm gg}(x)\!\!\! & \cong & \mbox{} \!\!
     + 2643.521\: {\cal D}_0 + 4425.451\: \delta(1-x)
     - 3590.1\: L_1 - 28489\: + 7469\: x 
  \nn \\[0.2mm] & & \mbox{}
     + 30421\: x^2
     - 53017\: x^3 + 19556\: x^4 - L_0 L_1\: ( 186.4 + 21328\: L_0 )
     + 12258\: L_0 
  \nn \\[0.2mm] & & \mbox{}
     + 13528\: L_0^2
     + 3281.7\: L_0^3 + 191.99\: L_0^4 + 5685.8\: x L_0^3
     + 14214.4\: x^{\,-1} 
  \nn \\[0.2mm] & & \mbox{}
     + 10233\: x^{\,-1} L_0 + 3651.1\: x^{\,-1} L_0^2
     + 3168\: x^{\,-1} L_0^3 + 576\: x^{\,-1} L_0^4
  \nn \\[0.7mm] &+& \mbox{} \nf \: \big( \:
     - 412.172\: {\cal D}_0 - 528.719\: \delta(1-x) + 319.97\: L_1
     + 248.95 + 260.6\: x 
  \nn \\[-0.5mm] & & \mbox{} \quad
     + 272.79\: x^2 + 2133.2\: x^3 - 926.87\: x^4
     + L_0 L_1\: ( 1266.5 - 29.709\: L_0 
  \nn \\[0.2mm] & & \mbox{} \quad
     + 87.771 L_1)
     + 4.9934\: L_0
     + 482.94\: L_0^2 + 155.1\: L_0^3 + 18.085\: L_0^4
  \nn \\[0.2mm] & & \mbox{} \quad
     + 485.18\: x L_0^3 
     - 804.13\: x^{\,-1} - 5.47\: x^{\,-1} L_0
     + 2368/9\: x^{\,-1} L_0^2 + 448/9\: x^{\,-1} L_0^3
     \,\big)
  \nn \\[0.7mm] &+& \mbox{} \n2f \: \big( \:
     - 16/9\: {\cal D}_0 + 6.4628\: \delta(1-x) - 77.19
     + 153.27\: x - 106.03\: x^2 + 11.995\: x^3
  \nn \\[-0.5mm] & & \mbox{} \quad
     - L_0 L_1\: ( 115.01 - 96.522\: x + 62.908\: L_0 )
     - 69.712\: L_0 - 44.8\: L_0^2 - 5.037\: L_0^3
  \nn \\[0.2mm] & & \mbox{} \quad
     + 472/243\: x^{\,-1} + 368/81\: x^{\,-1} L_0
     + 32/27\: x^{\,-1} L_0^2
     \,\big) \:\: .
\eea
In Eqs.~(\ref{eq:Pps-ap}) and (\ref{eq:Pgg-ap}) the small-$x$ leading terms are
exact up to truncations of irrational numbers. The same holds for the 
coefficients of $L_1$, $L_1^2$ and $L_1^3$ in Eq.~(\ref{eq:Pps-ap}) and that of
${\cal D}_0$ in Eq.~(\ref{eq:Pgg-ap}), where the coefficient of $\delta \x1$
has been minimally adjusted using the lowest moments.

The new NNLO off-diagonal quantities (\ref{eq:PTqg2}) and 
(\ref{eq:PTgq2}) can finally be parametrized as
\bea
\label{eq:Pqg-ap}
  P^{(2)T}_{\rm qg}(x)\!\!\! & \cong & \mbox{} \nf
     \big( \: 100/27\: L_1^4 + 350/9\: L_1^3 + 263.07\: L_1^2 
     + 693.84\: L_1 + 603.71 - 882.48\: x 
  \nn \\[-0.5mm] & & \mbox{} \quad     
     + 4723.2\: x^2
     - 4745.8\: x^3 - 175.28\: x^4 - L_0 L_1\, (1809.4 + 107.59\: x)
  \nn \\[0.2mm] & & \mbox{} \quad  
     - 885.5\: x L_0^4 + 1864\: L_0   
     + 1512\: L_0^2 + 361.28\: L_0^3 + 42.328\: L_0^4 
     + 1141.7\: x^{\,-1} 
  \nn \\[0.2mm] & & \mbox{} \quad     
     + 675.83\: x^{\,-1} L_0 
     - 64\: x^{\,-1} ( L_0^2 + L_0^3 ) \,\big) 
  \nn \\[0.7mm] &+& \mbox{} \!\!\! \n2f \, \big( 
     - 100/27\: L_1^3 - 35.446\: L_1^2 - 103.609\: L_1 
     - 113.81 + 341.26\: x - 853.35\: x^2 
  \nn \\[-0.5mm] & & \mbox{} \quad     
     + 492.1\: x^3 + 14.803\: x^4  
     + L_0 L_1\, ( 966.96  - 1.593\: L_1 - 709.1\: x ) 
     - 333.8\: xL_0^3 
  \nn \\[0.2mm] & & \mbox{} \quad     
     + 619.75\: L_0 + 255.62\: L_0^2 + 21.569\: L_0^3
     - 2.8986\: x^{\,-1} - 3.1752\: x^{\,-1} L_0 
  \nn \\[0.2mm] & & \mbox{} \quad     
     - 32/27\: x^{\,-1} L_0^2\, \big)
  \\[0.7mm] &+& \mbox{} \nft \: \big( \:
     4 + 6\: (L_0 + L_1) + ( 1 - 2\: x + 2\: x^2)\left(3.8696 
     + 4\: (L_0 + L_1) + 3\: (L_0 + L_1)^2\right) \big) \, \frak{4}{9}
\nn
\eea
where the $\nft$ part is exact, and

\pagebreak
\vspace*{-12mm}
\bea
\label{eq:Pgq-ap}
  P^{(2)T}_{\rm gq}(x)\!\!\! & \cong & \mbox{} 
     + 400/81\: L_1^4 + 520/27\: L_1^3 - 220.13\: L_1^2 
     - 152.6\: L_1 + 272.85 - 7188.7\: x 
  \nn \\[0.2mm] & & \mbox{} 
     + 5693.2\: x^2 
     + 146.98\: x^3 + 128.19\: x^4 - L_0 L_1 ( 1300.6 + 71.23 L_1 ) 
     + 543.8\: xL_0^3 
  \nn \\[0.2mm] & & \mbox{}      
     + 4.4136\: L_0 
     - 0.71252\: L_0^2 - 126.38\: L_0^3 - 30.061\: L_0^4 
     + 5803.7\: x^{\,-1} 
  \nn \\[0.2mm] & & \mbox{}      
     + 4776.5\: x^{\,-1} L_0 + 1001.89\: x^{\,-1} L_0^2 
     + 3712/3\: x^{\,-1} L_0^3 + 256\: x^{\,-1} L_0^4
  \nn \\[0.7mm] &+& \mbox{} \nf \: \big( \:
     80/81\: L_1^3 + 1040/81\: L_1^2 - 16.914\: L_1 - 871.3 
     + 790.13\: x - 241.23\: x^2 
  \nn \\[-0.5mm] & & \mbox{} \quad
     + 43.252\: x^3
     - 4.3465\: xL_0^3 + 55.048\: L_0 L_1 - 492\: L_0 
     - 343.1\: L_0^2 - 48.6\: L_0^3 
  \nn \\[0.2mm] & & \mbox{} \quad
     + 6.0041\: x^{\,-1} + 141.93\: x^{\,-1} L_0 
     + 2912/27\: x^{\,-1} L_0^2 + 1280/81\: x^{\,-1} L_0^3
     \,\big) \:\: .
\eea
The coefficients of the leading small-$x$ terms are exact in 
Eqs.~(\ref{eq:Pqg-ap}) and (\ref{eq:Pgq-ap}), up to a truncation of  
irrational numbers. The same holds for the coefficients of $L_0^2$, $L_0^3$ and
$L_0^4$ in Eq.~(\ref{eq:Pgq-ap}). The coefficients of the large-$x$ terms are 
also partially exact. 

Except for values of $x$ very close to zeros of the splitting functions, the 
parametrizations. (\ref{eq:P+appr}) -- (\ref{eq:Pgq-ap}) deviate from the exact 
expressions by less than one part in a thousand, which should be amply 
sufficient for foreseeable numerical applications.  
Also the complex-$N$ moments of the splitting functions can be readily obtained
to a perfectly sufficient accuracy using the above representations. 
The Mellin transform of Eqs.~(\ref{eq:P+appr}) -- (\ref{eq:Pgq-ap}) involve 
only simple harmonic sums (see, e.g, the appendix of 
Ref.~\cite{Blumlein:1998if}) of which the analytic continuations in terms
of logarithmic derivatives of Euler's $\Gamma$-function are well known.

\vspace{4mm}
{\small

\addtolength{\baselineskip}{-1.2mm}

}


\begin{thebibliography}{99}

\bibitem{PDG10}
K.~Nakamura et al.~[Particle Data Group],
 J. Phys. G37 (2010) 075021
 %%CITATION = JPHGB,G37,075021;%%

\bibitem{GP78}
H.~Georgi and H.D.~Politzer, 
 Nucl. Phys. B136 (1978) 445
 %%CITATION = NUPHA,B136,445;%%

\bibitem{AP77}
G. Altarelli and G. Parisi,
 Nucl. Phys. B126 (1977) 298
 %%CITATION = NUPHA,B126,298;%%

\bibitem{GL72}
V.N.~Gribov and L.N.~Lipatov,
 Sov. J. Nucl. Phys. 15 (1972) 438, ibid.~675
 %%CITATION = SJNCA,15,438;%%
 %%CITATION = SJNCA,15,675;%%

\bibitem{CFP80}
G.~Curci, W.~Furmanski and R.~Petronzio,
 Nucl. Phys. B175 (1980) 27
 %%CITATION = NUPHA,B175,27;%%

\bibitem{FP80}
W. Furmanski and R. Petronzio,
 Phys. Lett. 97B (1980) 437
 %%CITATION = PHLTA,97B,437;%%.

\bibitem{KKST80}
% J.~Kalinowski, K.~Konishi, (P.N. Scharbach) and T.~R.~Taylor
%J.~Kalinowski et al,
% Nucl. Phys. B181 (1981) 221, ibid.~253 
J.~Kalinowski, K.~Konishi, P.N. Scharbach and T.~R.~Taylor,
 Nucl. Phys. B181 (1981) 221, \\[1.8mm]
J.~Kalinowski, K.~Konishi and T.~R.~Taylor,
 Nucl. Phys. B181 (1981) 253
 %%CITATION = NUPHA,B181,221;%%
 %%CITATION = NUPHA,B181,253;%%

\bibitem{FKL81}
E.G.~Floratos, C.~Kounnas and R.~Lacaze,
 Nucl. Phys. B192 (1981) 417
 %%CITATION = NUPHA,B192,417;%%

\bibitem{MOKK01}
T.~Munehisa, H.~Okada, K.~Kudoh and K.~Kitani,
 Prog. Theor. Phys. 67 (1982) 609
 %%CITATION = PTPKA,67,609;%% 

\bibitem{MVV3}
S. Moch, J.A.M. Vermaseren and A. Vogt,
 Nucl. Phys. B688 (2004) 101, hep-ph/0403192
 %%CITATION = HEP-PH 0403192;%%.

\bibitem{MVV4}
A. Vogt, S. Moch and J.A.M. Vermaseren,
 Nucl. Phys. B691 (2004) 129, hep-ph/0404111 
 %%CITATION = HEP-PH 0404111;%%.

\bibitem{MMV06}
A.~Mitov, S.~Moch and A.~Vogt,
 Phys. Lett. B638 (2006) 61, hep-ph/0604053 
 %%CITATION = PHLTA,B638,61;%%

\bibitem{MV2}
S.~Moch and A.~Vogt,
  Phys. Lett.  B659 (2008) 290, arXiv:0709.3899 
  %%CITATION = PHLTA,B659,290;%%

\bibitem{DMS05}
Y.L.~Dokshitzer, G.~Marchesini and G.P.~Salam,
 Phys. Lett. B634 (2006) 504, hep-ph/0511302
 %%CITATION = HEP-PH 0511302;%%

\bibitem{BRvN00}
J.~Bl\"umlein, V.~Ravindran, W.L.~van Neerven,
 Nucl. Phys. B586 (2000) 349, hep-ph/0004172
 %%CITATION = NUPHA,B586,349;%%

\bibitem{SV96}
M.~Stratmann and W.~Vogelsang,
 Nucl. Phys. B496 (1997) 41, hep-ph/9612250
 %%CITATION = HEP-PH 9612250;%%

\bibitem{SMVV1}
G.~Soar, S.~Moch, J.A.M.~Vermaseren and A.~Vogt,
 Nucl. Phys.  B832 (2010) 152, arXiv:0912.0369
 %%CITATION = NUPHA,B832,152;%%

\bibitem{ZvN-F2}
E.B. Zijlstra and W.L. van Neerven,
 Phys.\ Lett.\ B272 (1991) 127,
 Phys.\ Lett.\ B273 (1991) 476
 %%CITATION = PHLTA,B272,127;%%.
 %%CITATION = PHLTA,B273,476;%%.

\bibitem{RvN96}
P.J.~Rijken and W.L.~van Neerven,
 Nucl.\ Phys.\ B488 (1997) 233, hep-ph/9609377
 %%CITATION = HEP-PH 9609377;%%

\bibitem{MV99}
S. Moch and J.A.M. Vermaseren,
  Nucl.\ Phys.\ B573 (2000) 853, hep-ph/9912355
  %%CITATION = NUPHA,B573,853;%%.

\bibitem{MM06}
A.~Mitov and S.~Moch,
  Nucl. Phys. B751 (2006) 18, hep-ph/0604160
  %%CITATION = NUPHA,B751,18;%%

\bibitem{DGGL10}
A. Daleo, A. Gehrmann-De Ridder, T. Gehr\-mann, G. Luisoni,
  JHEP 01 (2010) 118,~arXiv:0912.0374
  %%CITATION = JHEPA,1001,118;%%

\bibitem{BK93}
D.J. Broadhurst and A.L. Kataev,
  Phys. Lette B315 (1993) 179
  %%CITATION = PHLTA,B315,179;%%

\bibitem{Crewther}
R.J.~Crewther,
  Phys. Lett. B397 (1997) 137, hep-ph/9701321, \\[1.8mm]
  %%CITATION = PHLTA,B397,137;%%
%
%\bibitem{Baikov:2010je}
P.A.~Baikov, K.G.~Chetyrkin and J.H.~K\"uhn,
  Phys. Rev. Lett. 104 (2010) 132004,~arXiv:1001.3606
  %%CITATION = PRLTA,104,132004;%%

\bibitem{HPLs}
E.~Remiddi and J.A.M. Vermaseren,
 Int. J. Mod. Phys. A15 (2000) 725, hep-ph/9905237
 %%CITATION = HEP-PH 9905237;%%.

\bibitem{LLxto0}
A.H.~Mueller, Phys. Lett. B104 (1981) 161; \\[1.8mm]
 %%CITATION = PHLTA,B104,161;%%
A.~Bassetto, M.~Ciafaloni, G.~Marchesini and A.H.~Mueller,
%A.~Bassetto et al, 
 Nucl. Phys. B207 (1982) 189
 %%CITATION = NUPHA,B207,189;%%

\bibitem{AV2010}
A.~Vogt,
 Phys. Lett. B691 (2010) 77, arXiv:1005.1606 
 %%CITATION = PHLTA,B691,77;%%.

\bibitem{Grunb11}
G.~Grunberg, arXiv:1101.5377
 %%CITATION =  ARXIV:1101.5377;%%

\bibitem{BK06}
B.~Basso and G.P.~Korchemsky,
 Nucl. Phys. B775 (2007) 1, hep-th/0612247
 %%CITATION = NUPHA,B775,1;%%

\bibitem{AF81}
I.~Antoniadis and E.G.~Floratos,
 Nucl. Phys. B191 (1981) 217
 %%CITATION = NUPHA,B191,217;%%

\bibitem{MV5}
S.~Moch and A.~Vogt,
  JHEP 11 (2009) 099, arXiv:0909.2124
  %%CITATION = JHEPA,0911,099;%%

\bibitem{ASV1}
A.A.~Almasy, G.~Soar and A.~Vogt,
  JHEP 03 (2011) 030, arXiv:1012.3352
  %%CITATION = JHEPA,1103,030;%%

\bibitem{Hsums}
J.A.M. Vermaseren,
 Int. J. Mod. Phys. A14 (1999) 2037, hep-ph/9806280
 %%CITATION = IMPAE,A14,2037;%%.

\bibitem{HDecay}
K.G.~Chetyrkin, B.A.~Kniehl and M.~Steinhauser,
 Phys. Rev. Lett. 79 (1997) 353, hep-ph/9705240
 %%CITATION = PRLTA,79,353;%%

\bibitem{HDecay2}
M.~Schreck and M.~Steinhauser,
 Phys. Lett.  B655 (2007) 148, arXiv:0708.0916
 %%CITATION = PHLTA,B655,148;%%

\bibitem{Form3}
J.A.M. Vermaseren, {\it New features of FORM},
 math-ph/0010025
 %%CITATION = MATH-PH 0010025;%%.

\bibitem{HPLnum}
T. Gehrmann and E. Remiddi,
 Comput. Phys. Commun. 141 (2001) 296, hep-ph/0107173
 %%CITATION = HEP-PH 0107173;%%.
 
\bibitem{Blumlein:1998if}
J. Bl\"umlein and S. Kurth,
  Phys. Rev. D60 (1999) 014018, hep-ph/9810241
  %%CITATION = PHRVA,D60,014018;%%. 

\end{thebibliography}
\end{document}